\documentclass[aip,jcp,reprint]{revtex4-1}

\usepackage{amsmath}
\usepackage{amssymb}
\usepackage{graphicx}
\usepackage{dcolumn}
\usepackage{bm}
\usepackage{amsthm}

\usepackage[utf8]{inputenc}
\usepackage[T1]{fontenc}
\usepackage{mathptmx}
\usepackage{etoolbox}

\usepackage{lipsum}
\usepackage[ruled, vlined, linesnumbered]{algorithm2e}
\SetKwProg{custom}{}{}{}
\usepackage{xcolor}
\usepackage{subcaption}
\usepackage{soul}
\usepackage{color}

\usepackage{ulem}
\newcommand\Lukesout{\bgroup\markoverwith{\textcolor{blue}{\rule[0.5ex]{2pt}{0.4pt}}}\ULon}
\usepackage{hyperref}
\hypersetup{hidelinks, 
}
\newtheorem{theorem}{Theorem}

\newtheorem{lemma}[theorem]{Lemma}

\newtheorem{assumption}{Assumption}

\newtheorem*{notation}{Notation}

\newcommand{\R}{\mathbb R}

\DeclareMathOperator*{\argmax}{argmax}

\DeclareMathOperator{\tr}{tr}

\let\oldnl\nl
\newcommand{\nonl}{\renewcommand{\nl}{\let\nl\oldnl}}

\renewcommand{\thesection}{\Roman{section}}
\renewcommand{\thesubsection}{\Alph{subsection}}
\renewcommand{\thesubsubsection}{\arabic{subsubsection}}
\makeatletter
\renewcommand{\p@subsection}{\thesection.}
\renewcommand{\p@subsubsection}{\thesection.\thesubsection.}
\makeatother

\makeatletter
\def\@email#1#2{%
 \endgroup
 \patchcmd{\titleblock@produce}
  {\frontmatter@RRAPformat}
  {\frontmatter@RRAPformat{\produce@RRAP{*#1\href{mailto:#2}{#2}}}\frontmatter@RRAPformat}
  {}{}
}%
\makeatother
\begin{document}

\preprint{AIP/123-QED}

\title[Computing committors via Mahalanobis diffusion maps with enhanced sampling data]{Computing committors via Mahalanobis diffusion maps with enhanced sampling data}
\author{L. Evans}
\homepage{https://www.math.umd.edu/~evansal/}
\email{evansal@umd.edu}
\author{M.K. Cameron}%
\homepage{http://www.math.umd.edu/~mariakc/}
\email{mariakc@umd.edu}
\affiliation{Department of Mathematics, University of Maryland, College Park, MD 20742, USA
}%
\author{P. Tiwary}
\homepage{https://go.umd.edu/tiwarylab}
\email{ptiwary@umd.edu}

\affiliation{%
  Department of Chemistry and Biochemistry and Institute for Physical Science and Technology, University of Maryland, College Park, MD 20742, USA
}%

\date{\today}

\begin{abstract}
The study of phenomena such as protein folding and conformational changes in molecules is a central theme in chemical physics. Molecular dynamics (MD) simulation is the primary tool for the study of transition processes in biomolecules, but it is hampered by a huge timescale gap between the processes of interest and atomic vibrations which dictate the time step size. Therefore, it is imperative to combine MD simulations with other techniques in order to quantify the transition processes taking place on large timescales. In this work, the diffusion map with Mahalanobis kernel, a meshless  approach for approximating the Backward Kolmogorov Operator (BKO) in collective variables, is upgraded to incorporate standard enhanced sampling techniques such as metadynamics. The resulting algorithm, which we call the \emph{target measure Mahalanobis diffusion map} ({\tt tm-mmap}), is suitable for a moderate number of collective variables in which one can approximate the diffusion tensor and free energy.
Imposing appropriate boundary conditions allows use of the approximated BKO to solve for the committor function and utilization of transition path theory  to find the reactive current delineating the transition channels and the transition rate. 
The proposed algorithm, {\tt tm-mmap}, is tested on the two-dimensional Moro-Cardin two-well system with position-dependent diffusion coefficient and on alanine dipeptide in two collective variables where the committor, the reactive current, and the transition rate are compared to those computed by the finite element method (FEM). Finally, {\tt tm-mmap} is applied to alanine dipeptide in four collective variables where the use of finite elements is infeasible.
\end{abstract}

\maketitle

\section{\label{sec:intr}Introduction}
Molecular dynamics (MD) simulations provide an atomistic-resolution lens for probing chemical systems. These systems commonly have high dimensionality, reside in metastable states on extremely large time scales, and undergo transitions on extremely small time scales relative to residence times in the metastable states. These transitions, such as protein folding or
conformational changes in a molecule, are crucial to molecular simulations
but difficult to characterize due to the timescale gap, and are thus commonly referred to as rare events. The transition path theory (TPT)~\cite{weinan2004, weinan2006towards,EVE2011} is a mathematical framework for the direct study of such rare transitions in stochastic systems, and it is particularly utilized for metastable systems arising in MD. TPT is framed around the committor function, an optimal reaction coordinate with which reaction rates and reaction channels can be computed between a predefined reactant state $A$ and product state $B$. In practice, the committor is very difficult to compute globally: statistical or simulation-based methods require observing large numbers of rare transitions while mathematical partial differential equation (PDE)-based approaches require  gridding the state space and thus are limited to low dimensions.
Instead, numerical TPT for high-dimensional systems typically involves finding a zero-temperature asymptotic transition path and using umbrella sampling or related techniques to access the committor.~\cite{weinan2006towards}
This approach is viable but assumes that rare transitions are localized to a narrow tube around the found path. In practice, transition processes can be broad and complex~\cite{satija2020broad} and the computed path may misrepresent the true kinetics.

For a more global analysis one can instead utilize collective variables to lower the dimensionality and increase interpretability of existing methods.~\cite{cameron2013estimation, 2006string}
However, a biomolecule may still require a large number of collective variables such as contact distances or dihedral angles for adequate representation .~\cite{ravindra2020automatic,rohrdanz2011determination, piana2008advillin,sittel2018perspective}
Hence, one still may be unable to leverage traditional mesh-based PDE solvers to even reduced dimensionality data given either in physics-informed or machine-learned collective variables.

In contrast to traditional PDE solvers, meshless approaches to computing the committor discretize the backward Kolmogorov PDE to a point cloud obtained from a molecular dynamics simulation. Recent approaches of this type include parametrizing the committor with a neural network~\cite{rotskoff2020,li2021semigroup,ren2019,khoo2019solving} and approximating dynamical operators with diffusion maps.~\cite{trstanova2020local, evans2021computing} 

The diffusion map algorithm introduced by Coifman and Lafon~\cite{coifman2006diffusion} in 2006 is a widely used dimension reduction algorithm which provably approximates differential operators on unstructured datasets. Its extensions most relevant to the present work include several works concerning the sampling and transformation of the input data. The Mahalanobis diffusion map \cite{singer2008} untangles the effect of nonlinear transformation applied to the data.  Integration with umbrella sampling \cite{ferguson2011integrating} uncovers informative low-dimensional local parametrizations for the high-dimensional system of interest. Local kernels \cite{berry2016local} adapt diffusion maps for data coming from processes governed by arbitrary Ito SDEs. The \emph{target measure diffusion map} \cite{banisch2020,trstanova2020local} allows for the data to be sampled from an arbitrary density provided that the invariant density for the process of interest is known.

The original diffusion map algorithm \cite{coifman2006diffusion} relies on the assumption that the input data are sampled from a Gibbs distribution of interest and that the underlying process is governed by the overdamped Langevin dynamics.
{ In our recent work \cite{evans2021computing}, the diffusion map algorithm was promoted to approximate the dynamics in collective variables that is time-reversible but { typically has} a position-dependent and anisotropic diffusion tensor $M(x)$. The effect of such a diffusion was captured via the use of the Mahalanobis kernel~\cite{singer2008} 
$$
k_{\epsilon}^{\tt mmap}(x,x') = \exp\left\{-\frac{(x-x')^\top [ M(x)+M(x')](x-x')}{4\epsilon}
\right\}
$$
and required the calculation of the diffusion tensor in collective variables at each data point by umbrella sampling. \cite{2006string} We proved that the resulting Mahalanobis diffusion map ({\tt mmap}) accurately approximates the generator in collective variables and numerically validated the algorithm on simple MD systems.~\cite{evans2021computing}}

{ A substantial practical shortcoming of {\tt mmap} was that it required the input data sampled from the invariant density, while often an adequate dataset for the study of a transition of interest can be achieved only with enhanced sampling. A vivid illustration for it on the example of alanine dipeptide is given in Ref.~\onlinecite{ren2019}.}

In this work, we extend {\tt mmap}~\cite{evans2021computing} to work with input data generated by any standard enhanced sampling technique. {This is achieved by the following right-normalization of the Mahalanobis kernel:
$$
k_{\epsilon,\mu}^{\tt tm-mmap}(x,x') 
=k_{\epsilon}^{\tt mmap}(x,x')\frac{\mu^{1/2}(x')|M(x')|^{-1/4}}{\rho_{\epsilon}(x')}.
$$
Here, $\mu$ is the target measure, i.e., the invariant measure for the dynamics in collective variables, $|M|$ is the determinant of the diffusion tensor $M$, and $\rho_{\epsilon}$ is an estimate for the sampling density $\rho$ that can be readily obtained in the diffusion map framework.}
This renormalization is based on a reweighting scheme for the rotationally symmetric Gaussian kernel 
proposed by Banisch et al.~\cite{banisch2020,trstanova2020local}.
We will refer to the resulting algorithm as the \emph{target measure Mahalanobis diffusion map} and abbreviate it as {\tt tm-mmap}.

It is important to note that for metastable molecular systems, quantities like the committor are relevant precisely where sampling is extremely difficult. Enhanced sampling approaches such as umbrella sampling~\cite{torrie1977nonphysical} and metadynamics~\cite{laio2002escaping} can alleviate this issue. The proposed algorithm, {\tt tm-mmap}, combines {\tt mmap} with the enhanced sampling. Its output is the committor function that is processed via the transition path theory to obtain the reactive current and the transition rate between the metastable states of interest. In summary, {\tt tm-mmap} is a simple and customizable tool which expands the diffusion map framework to { a moderate number of user-defined collective variables and enhanced} sampling for rare event analysis in molecular dynamics. 

We validate {\tt tm-mmap} on the Moro-Cardin two-well system with position-dependent diffusion~\cite{moro1998saddle} and on alanine dipeptide with two dihedral angles in vacuum. Here metadynamics~\cite{laio2002escaping} is used for enhanced sampling, though our method could easily work in conjunction with other enhanced sampling approaches. We compare the {\tt tm-mmap} committor, the reactive current, and the transition rate to those computed using the finite element method and show that the {\tt tm-mmap} committors and rates are robust with respect to the choice of the kernel bandwidth. 
{ Importantly, the kernel bandwidth values chosen from a standard heuristic \cite{berry2015nonparametric,berry2016variable,giannakis2019data,davis2021graph} lead to near-optimal error for the committor on a variety of subsampled datasets, in particular for spatially quasi-uniformly data.}

Finally, we apply {\tt tm-mmap} to alanine dipeptide with four dihedral angles in vacuum and demonstrate its ability to facilitate the quantification of the transition process between the C7eq and C7ax metastable states in 4D. Moreover, the transition rate computed for this transition using {\tt tm-mmap} is in good agreement with the estimate obtained by simulating a very long unbiased trajectory in Ref.~\onlinecite{vani2022computing}.

The rest of the paper is organized as follows. Background on the Langevin dynamics, collective variables, the transition path theory, and diffusion maps is given in Section \ref{sec:background}. The proposed methodology including the {\tt tm-mmap} algorithm as well as algorithms for choosing parameters and processing input data are detailed in Section \ref{sec:methodology}. Numerical tests are presented in Section \ref{sec:examples}. An application to alanine dipeptide in four dihedral angles is reported in Section \ref{sec:aladip4D}. A discussion of the results is found in Section \ref{sec:discussion}. Conclusions are summarized in Section \ref{sec:conclusion}. A convergence proof for {\tt tm-mmap} is placed in Appendix \ref{sec:appendixA}.

\section{Background}
\label{sec:background}

\subsection{\label{sec:colvars} Effective dynamics in collective variables}

Our primary interest in this work is in datasets arising in MD simulations, though we believe the formalism developed here would be applicable to generic trajectories across chemical physics obtained from simulations or experiments.
We consider the 
\emph{Langevin dynamics}, a commonly used model for molecular motion which describes the molecular configuration in terms of the atomic positions $y$ and their velocities $v$:
\begin{align}
dy & = v dt\label{eqn:langevin}\\
dv & = -\left(m^{-1}\nabla V(y)+\gamma v\right)dt + \sqrt{2\gamma(m\beta)^{-1}}dw.
\end{align}
Here $V:\R^{n} \to \R$ is a potential
function, $\beta^{-1} = k_b T$ is temperature in units of
Boltzmann's constant, $t$ is time, $\gamma$ is the friction coefficient, $m$ is the diagonal $n\times n$ mass matrix, and $w_t$ is a Brownian motion in
$\R^{n}$. 
We assume that the potential $V(y)$ is such that the system
governed by \eqref{eqn:langevin} is ergodic with respect to the invariant Boltzmann-Gibbs density  
{
$$
\mu(y,v) = \frac{\beta^{n/2}|m|^{1/2}}{Z_V(2\pi)^{1/2}}
\exp\left\{-\beta\left( V(y)+\tfrac{1}{2}v^\top mv\right)\right\},
$$ 
where $|m|$ denotes the determinant of the mass matrix $m$ and $Z_{V} = \int_{\mathbb{R}^n}\exp(-\beta V(y))dy$.
}

As mentioned in Introduction, the number of atoms in biomolecules is typically very large. Even the small molecule alanine dipeptide comprises 22 atoms, and hence has a 66-dimensional configuration space ($n=66$) and 132-dimensional phase phase $(y,v)$.
On the other hand, an effective description of the state of a biomolecule is usually done in terms of certain functions in $y$
specifying desired geometric characteristics rather than in terms of $(y,v)$.
Therefore, to reduce the dimensionality and obtain a more useful and comprehensive description of the system-at-hand, one uses
\emph{collective variables} (CVs).
CVs are functions of the atomic coordinates
designed to give a coarse-grained description of the system's dynamics,
preserving transitions between metastable states but erasing small-scale
vibrations. Physical intuition has traditionally driven the choice of
collective variables including
dihedral angles, intermolecular distances, macromolecular distances, and
various experimental measurements.

For a given system configuration $y\in \R^n,$
we denote the CV coordinates as $\theta(y) = (\theta_1(y), \ldots, \theta_{d}(y))^{\top} \in\Omega$ where $d \ll n$ and $\Omega$ is a $d$-dimensional manifold. { Since the CVs are dihedral angles or distances, we limit our considerations to the case where $\Omega$ is of the form $\mathbb{T}^k\times\mathbb{R}^{d-k}$, where $k$ can be $0,1,\ldots,d$. I.e., the manifold $\Omega$ is a direct product of tori and real lines and its local geometry coincides with the one in $\mathbb{R}^d.$}


We assume that $\theta(y)$ is a sufficiently ``good' set of CVs (given more precisely in Section~\ref{sec:tpt}) such that we can describe a closed dynamics in coordinates $x = \theta(y)$ by the anisotropic overdamped Langevin dynamics of the form~{\cite{2006string}}
\begin{align}
  dx_t &= \left(-M(x_t)\nabla F(x_t) + \beta^{-1}\nabla \cdot M(x_t)\right)dt\notag\\
  & + \sqrt{2\beta^{-1}}M^{1/2}(x_t) dW_t,  \label{eqn:cvsde}
\end{align}
 where
$F(x)$ is the \emph{free energy} with respect to the CVs given by
\begin{align}
  \label{eqn:free_energy_formula}
  F(x) = - \beta^{-1} \ln \left(\int_{\R^{n}} Z_V^{-1} e^{-\beta V(y)} 
  \delta(\theta(y) - x) dy\right),
\end{align}
and $M(x)$ is the \emph{diffusion matrix} for the CVs defined by
\begin{align}
  \label{eqn:diffusion_tensor_formula}
  M(x) = e^{\beta F(x)}\int_{\R^{n}} J(y)m J^{\top}(y) Z_{V}^{-1} e^{-\beta V(y)}
  \delta(\theta(y) - x) dy.
\end{align}
Here $J(y)$ is the $d\times n$ Jacobian matrix with entries
$$
  J_{ij}(y) = \frac{\partial\theta_i(y)}{\partial y_j} \qquad 1 \le i \le d, \quad 1 \le j \le n.
$$
Note that since the collective variables depend only on the positions $y$,  the velocities $v$ are integrated out in deriving the effective dynamics~\eqref{eqn:cvsde}.
{The practical calculation of $M(x)$ uses the averaging of $JJ^{\top}$ over the data generated using restrained dynamics, as detailed in Ref.~\onlinecite{2006string} and Appendix A of Ref.~\onlinecite{evans2021computing}.}


One can check that the invariant probability measure for the process governed by SDE \eqref{eqn:cvsde} is the Gibbs density
\begin{equation}
\label{eqn:Gibbs_cv}
\rho(x)=Z_F^{-1} e^{-\beta F(x)},
\end{equation}
and that the process is time-reversible~\cite{weinan2006towards}. Moreover, any reversible diffusion process must be of the form \eqref{eqn:cvsde}  for some function $F(x)$ and some matrix $\beta^{-1}M(x)$.  The drift term $\beta^{-1}\nabla \cdot M(x)$ is the correction that ensures reversibility.~\cite{pavliotis2014stochastic} It is a column vector whose $i$th component is 
$$
  \beta^{-1}[\nabla \cdot M(x)]_i = \beta^{-1}\sum_{j=1}^{d} \frac{\partial M_{ij}(x)}{\partial x_j} \qquad 1 \le i \le d.
$$

For the datasets considered in this work, we compute the diffusion matrix $M(x)$ via local simulations with quadratic position restraints as described in Ref. \onlinecite{2006string}.
This procedure is an outgrowth of well-established uses for restrained dynamics within the molecular
dynamics community, particularly in
position-dependent friction~\cite{straub1987calculation} as well as
fundamental works for computing free
energy differences.~\cite{carter1989constrained,kirkwood1935statistical}

The generator for the SDE \eqref{eqn:cvsde} is given by
\begin{equation}
  \label{eqn:mgen}
  \mathcal{L}f = \left[-M\nabla F + \beta^{-1}(\nabla \cdot M)\right]^{\top}\nabla f + \beta^{-1}{\sf tr}[M\nabla \nabla f],
\end{equation}
which can also be written in divergence form as
\begin{equation}\label{eqn:mgen_divergence}
  \mathcal{L}f = \beta^{-1} e^{\beta F} \nabla \cdot(e^{-\beta F}M \nabla f).
\end{equation}

{ We would like to remark that we have chosen model \eqref{eqn:cvsde} for the dynamics in collective variables due to the fact that it is time-reversible and amenable for the diffusion maps framework. There are other models for the dynamics in collective variables rigorously derived in Refs.~\onlinecite{zhang2017effective} and~\onlinecite{legoll2010effective} that have, in general, a different drift from \eqref{eqn:cvsde} and match \eqref{eqn:cvsde} if the underlying system evolves according to the overdamped Langevin dynamics.}

\subsection{\label{sec:tpt} Transition Path Theory}
Let $\{x_t\}_{t=0}^{\infty}$ be an infinite trajectory of a system evolving according to SDE \eqref{eqn:cvsde}. We are interested in transitions between disjoint neighborhoods $A$ and $B$ of minima $x_A$ and $x_B$ of the potential function $F(x)$. The sets $A$ and $B$ are referred to as the \emph{reactant} and
\emph{product} sets respectively.
\begin{figure}[t]
  \begin{center}
    \includegraphics[width=0.5\textwidth]{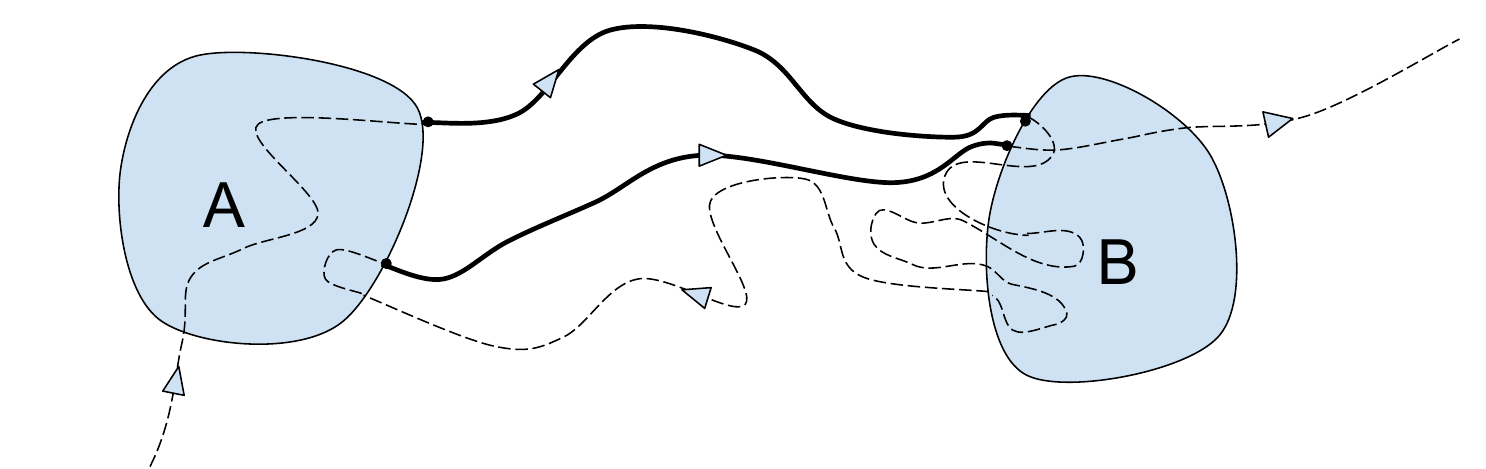}
    \caption{
      {\small  A segment of a long trajectory. Reactive pieces from reactant state $A$ to product state $B$ are shown with solid lines.}
    }
    \label{fig:reactive}
  \end{center}
\end{figure}
Transition path theory (TPT)~\cite{weinan2006towards,EVE2011} is a mathematical framework to analyze statistics of transitions
between the reactant $A$ and the product $B.$
The subject of TPT is the ensemble of \emph{reactive trajectories}, i.e.
continuous pieces of the trajectory $x_t$ which start at $\partial A$ and
end at $\partial B$ without returning to $\partial A$ in-between (see
Figure~\ref{fig:reactive}).
Key concepts of TPT are the
\emph{forward committor} and \emph{backward committor} functions with
respect to $A$ and $B$. Since the governing SDE \eqref{eqn:cvsde} is reversible,
the forward $q_{+}$ and backward $q_{-}$ committors are related via \cite{weinan2006towards}  $q_{-}=1-q_{+}$.
Hence, for brevity, we will refer to the forward committor as \emph{the committor} and denote it by $q(x)$.
The committor $q$ has a straightforward probabilistic interpretation:
\begin{equation}
  \label{eqn:comm_def}
  q(x) = \mathbb{P}(\tau_B < \tau_{A} \mid x_0=x),
\end{equation}
where $\tau_A:= \inf \lbrace t > 0 \mid x_t \in A \rbrace$ and $ \tau_B = \inf \lbrace t > 0 \mid x_t \in B \rbrace$ are
the \emph{first entrance times} of the sets $A$ and $B$ respectively.
In words, $q(x)$ is the probability that a trajectory starting at $x$ will arrive at the product set $B$ before arriving at the reactant set $A$.

One can show that $q$ satisfies the boundary value problem~\cite{weinan2006towards}
\begin{equation}\label{eqn:comm_pde}
  \begin{cases}
    \mathcal{L}q(x) =  0 & x \in \Omega \backslash (A \cup B), \\
    q(x) = 0             & x \in \partial A,                   \\
    q(x) = 1             & x \in \partial B,                  \\
  \end{cases}
\end{equation}
where $\mathcal{L}$ is the infinitesimal generator from equation~\eqref{eqn:mgen}.

Although the committor can be considered an optimal reaction coordinate,~\cite{ma2006dynamic,peters2016reaction} the committor by itself does not illustrate the mechanism of the transition process. Instead, from the committor one can compute the \emph{reactive current}, a vector field defined by
\begin{equation}
  \label{eqn:cvrcur}
  \mathcal{J}(x) = \beta^{-1}Z_F^{-1}e^{-\beta F(x)}M(x)\nabla q(x).
\end{equation}
The reactive current gives a complete quantitative characterization of the transition process from $A$ to $B$. Using it, one can identify reactive channels and compute the reaction rates.
The integral of the flux of the reactive current through any hypersurface $\Sigma$
separating the sets $A$ and $B$ gives the reaction rate:
\begin{equation}
  \label{eqn:nuAB}
  \nu_{AB} :=\lim_{t\rightarrow\infty} \frac{N_{AB}}{t}= \int_{\Sigma}\mathcal{J}\cdot{\hat{n}}d\sigma,
\end{equation}
where $N_{AB}$ is the total number of transitions from $A$ to $B$ performed by the system
within the time interval $[0,t]$, and $\hat{n}$ is the unit normal to the surface $\Sigma$ pointing in the direction of $B$. {For the reactive current given by \eqref{eqn:cvrcur} it can be shown by choosing $\Sigma$ to be an isocommittor surface that the transition rate can be found as
\begin{equation}
\label{eqn:nuAB1}
\nu_{AB} = \beta^{-1}Z_F^{-1}\int_{\Omega\backslash(A\cup B)}e^{-\beta F}\nabla q^\top M\nabla q dx. 
\end{equation}
}

Another quantity of interest is the probability density of reactive trajectories $\pi_{AB}(x),$ defined by
\[\pi_{AB}(x) = Z_F^{-1}e^{-\beta F(x)}q(x)(1 - q(x)), \]
which is the probability density to observe a reactive trajectory at $x \notin (A \cup B)$ at time $t.$ Note that 
\begin{equation}
\label{eqn:rhoAB}
\rho_{AB}:=\int_{\Omega}\pi_{AB}dx
\end{equation}
is the probability that an infinite trajectory $x_t$ is reactive at any randomly picked moment of time $t$.

Since our goal is to compute the committor and related quantities, 
our criterion {for a set of good CVs} is  determined by how well the committor for SDE \eqref{eqn:cvsde} approximates the committor for the Langevin dynamics \eqref{eqn:langevin} 
{ in the region of the phase space $(y, v)$ where the density of reactive trajectories is substantial}.
Hence, we say that a set of CVs $y$ is good if the integral
\begin{equation}
\label{eqn:cverr}
\int_{\mathbb{R}^{2n}} |q(\theta(y)) - q(y,v)| \tilde{\pi}_{\tilde{A}\tilde{B}}(y,v)dydv,
\end{equation}
is small, where the ``lifted'' reactant and product sets are defined as $\tilde{A}:=\{(y,v)~|~\theta(y)\in A, \|v\| < r\}$, $\tilde{B}$ likewise, and  $\tilde{\pi}_{AB}(y,v)$ is the probability density of reactive trajectories in the original variables $(y,v)$. This criterion is essentially used when the committor is checked by \emph{committor analysis}.~\cite{geissler1999kinetic} 


\subsection{\label{sec:level2}Diffusion maps}

%
%
Diffusion maps, like spectral clustering~\cite{von2007tutorial} and Laplacian eigenmaps,~\cite{belkin2003laplacian} impose an affinity or similarity measure on a dataset defined by a kernel function, and then analyze the associated graph on the dataset. One interpretation of diffusion maps is that a particular Markov Chain is defined on the data, and analysis of the Markov chain reveals additional structure about the data.
An important difference between diffusion maps and previous algorithms, in particular Laplacian eigenmaps, \cite{belkin2003laplacian} is the emphasis on the discrete-to-continuous interplay of the kernel matrix and algorithmic utilization of known asymptotic error results.
Perhaps most important is that the random-walk graph Laplacian on a dataset with sampling density $\rho(x)$ corresponds to an overdamped Langevin dynamics with stationary measure $\rho^2(x)$ instead of $\rho(x).$ Thus, for non-uniform densities such as a Gibbs distribution, graph Laplacian-like methods must account for the sampling density to obtain correct estimates. Coifman and Lafon~\cite{coifman2006diffusion} in introducing diffusion maps tune the sampling so that for any choice of scalar parameter $\alpha$ the diffusion map corresponds to a gradient dynamics with stationary density $\rho^{2(1 - \alpha)}.$ 

The diffusion map algorithm takes as input a dataset $X = \{x_i\}_{i=1}^N \subset \R^d$ with sampling density $\rho(x).$ For application to molecular dynamics, $x_i$ is usually a vector representing the $i$-th molecular conformation from a trajectory, but can generally be considered as a $d-$dimensional feature vector for the $i$-th piece of data. Pairwise similarity of data is encoded through a distance function $d(x,x')$ and kernel function $k_{\epsilon}(x,x'),$ whose simplest form is given by
\begin{equation}\label{eqn:gaussian_kernel}
  k_{\epsilon}(x,x') = \exp\left(-\frac{d(x,x')^2}{2\epsilon}\right).
\end{equation}
For many applications $d(x,x')$ can be taken as Euclidean distance between the vectors $x,x'$, but many other distances can be used.~\cite{tsai2021distance,boninsegna2015investigating} For datasets consisting of molecular configurations in atomic coordinates, the most commonly used distance is the root-mean-square deviation (RMSD) between configurations.~\cite{ferguson2010systematic,zheng2011polymer} Other distances include Euclidean distance after RMSD-aligning configurations to a reference structure~\cite{coifman2008diffusion,trstanova2020local} and hybrid RMSD-energy based distances,~\cite{tan2019approximating} with others compared in Ref.~\onlinecite{boninsegna2015investigating} as structural and kinetic similarity measures.~\cite{tsai2021sgoop}

The user-chosen parameter $\epsilon > 0$ is the \emph{kernel bandwidth}. The kernel defines an $N \times N$ similarity matrix $K_{\epsilon}$ with $[K_{\epsilon}]_{ij} = k_{\epsilon}(x_i, x_j).$
We can then consider a graph on $i$ nodes with feature vectors $X$ and weighted adjacency matrix given by $K_{\epsilon}.$ Our goal is to use the discrete graph defined by our data $X$ and the kernel matrix $K_{\epsilon}$ to analyze the continuous process generating the data.

The discrete-to-continuous interplay arises from Monte Carlo integration: 
%
for an observable $f(x)$ and for a sufficiently large dataset with sampling density $\rho(x),$
$\frac{1}{N}[K_{\epsilon}f]_i$ is the Monte Carlo approximation for the average of $k_{\epsilon}(x_i,\cdot)f(\cdot)$ with respect to $\rho(x):$
\begin{equation}
  \label{eqn:montecarlo}
  \lim \limits_{n \to \infty}\frac{1}{N} \sum\limits_{j = 1}^{N}
  k_{\epsilon} (x_i, x_j) f(x_j) = \int_{\Omega} k_{\epsilon} (x_i, x') f(x')
  \rho(x') dx'
\end{equation}
with error decaying as $O(N^{-\frac{1}{2}})$.



\subsubsection{Density estimation}
A natural interpretation of the Monte Carlo integration with the kernel $k_{\epsilon}(x,y)$ given by~\eqref{eqn:gaussian_kernel} suggests that, for small $\epsilon$,
\begin{equation}
\label{eqn:deltaapprox}
\frac{1}{c_{\epsilon}(x)}k_{\epsilon}(x,x') \approx \delta(x - x') 
\end{equation}  
where  $c_{\epsilon}(x)$ is the normalizing factor for the kernel $k_{\epsilon}(x,x')$ defined by
\begin{equation}
\label{eqn:ce}
c_{\epsilon}(x):=\int_{\Omega}k_{\epsilon}(x,x')dx'.
\end{equation}

%
{Motivated by \eqref{eqn:deltaapprox}, we define the
continuous \emph{kernel density estimator} as
\begin{equation}\label{eqn:mmap_density_original}
\rho_{\epsilon}(x):= \frac{1}{c_{\epsilon}(x)}\int_{\Omega} k_{\epsilon}(x, y) \rho(y) dy
\end{equation}
and its discrete counterpart as the vector
$p_{\epsilon} \in \R^{N}$ with
\begin{equation}\label{eqn:kde}
  [p_{\epsilon}]_i = \frac{1}{N c_{\epsilon}(x_i)}\sum\limits_{j=1}^N k_{\epsilon}(x_i, x_j).
\end{equation}}
%


\subsubsection{Approximating differential operators}
\label{sec:diffoper}
A crucial use of the kernel function $k_{\epsilon}(x,y)$ in diffusion maps is for computing { actions of certain differential operators on functions} 
without gridding the phase space. {For example, the standard Gaussian kernel 
\begin{equation}
\label{eqn:gausskernel}
k_{\epsilon}(x,x') := e^{-\frac{||x - x'||^2}{2\epsilon}}
\end{equation} 
on $\Omega = \R^d$ is the heat kernel for the backward Kolmogorov equation for the standard Brownian motion: $\frac{ \partial u}{\partial \epsilon} = 0.5\Delta u$.  The normalizing factor for this kernel is $c_{\epsilon} = (2\pi \epsilon)^{d/2}$. The parameter $\epsilon$ plays the role of time. Given an initial condition $u(0,x) = f(x)$ at $\epsilon = 0$, one can find $u(\epsilon,t)$  at any $\epsilon>0$:
$$
u(\epsilon,x) = \frac{1}{(2\pi \epsilon)^{d/2}}\int_{\R^d} e^{-\frac{||x - x'||^2}{2\epsilon}} f(x') dx'.
$$
On the other hand, for small $\epsilon$, $u(\epsilon,x)$ can be found using the Taylor expansion:
\begin{align*}
u(\epsilon,x) &= u(0,x) + \frac{\partial}{\partial \epsilon}u(0,x) + O(\epsilon^2)  \\
& = f(x) + \frac{\epsilon}{2}\Delta f(x) + O(\epsilon^2).
\end{align*}
Therefore, the following identity must hold:
\begin{equation}
\label{eqn:gaussian_expansion}
  \frac{1}{c_{\epsilon}}\int_{\R^d} e^{-\frac{||x - x'||^2}{2\epsilon}} f(x') dx'  = f(x) + \frac{\epsilon}{2}\Delta f(x) + O(\epsilon^2).
\end{equation}

Now we observe that
$$
\lim_{N\rightarrow\infty}\frac{1}{N}\sum_j e^{-\frac{||x - x_j||^2}{2\epsilon }}f(x_j) =
\int_{\R^d} e^{-\frac{||x - x'||^2}{2\epsilon}} f(x')\rho(x') dx',
$$
where $\rho$ is the sampling density. Therefore,
\begin{align}
&\lim_{N\rightarrow\infty}
\frac{\sum_j  e^{-\frac{||x - x_j||^2}{2\epsilon }}f(x_j)}
{\sum_j  e^{-\frac{||x - x_j||^2}{2\epsilon }}} = 
\frac{\int_{\R^d} e^{-\frac{||x - x'||^2}{2\epsilon}} f(x')\rho(x) dx'}{
\int_{\R^d} e^{-\frac{||x - x'||^2}{2\epsilon}} \rho(x) dx'
} \label{eqn:Markov1}\\
=& \frac{(f\rho)(x) + \frac{\epsilon}{2}\Delta( f\rho)(x) + O(\epsilon^2)}
{\rho(x) + \frac{\epsilon}{2}\Delta \rho(x) + O(\epsilon^2)} \notag\\
=& f(x) +\frac{\epsilon}{2}\left[ \Delta f(x) + 2\nabla f(x)\cdot\frac{\nabla\rho(x)}{\rho(x)}\right] + O(\epsilon^2).\label{eqn:Markov2}
\end{align}
Choosing $x = x_i$, $i=1,\ldots,N$, we observe that the fraction in the left-hand side of \eqref{eqn:Markov1} can be viewed as a Markov matrix $P_{\epsilon}$ with entries
$$
[P_{\epsilon}]_{ij} = 
\left[\sum_l  e^{-\frac{||x_i - x_l||^2}{2\epsilon }}\right]^{-1}
 e^{-\frac{||x_i - x_j||^2}{2\epsilon }}
$$
multiplied by the vector $[f(x_j)]_{j=1}^N$, and its limit, i.e., the fraction in the right-hand side of \eqref{eqn:Markov1}, is the corresponding Markov operator $\mathcal{P}_{\epsilon}$  acting on $f(x)$. Equation \eqref{eqn:Markov2} implies that 
\begin{equation}
\label{eqn:gensimple}
\lim_{\epsilon\rightarrow 0}\frac{\mathcal{P}_{\epsilon}f - f}{\epsilon} = \Delta f + \nabla f\cdot\nabla\left[\log\rho^2\right],
\end{equation}
which is the generator for the overdamped Langevin dynamics with  a temperature twice as high as the  sampling density corresponds to -- exactly as 
mentioned in Section \ref{sec:level2}. 
}

Coifman and Lafon~\cite{coifman2006diffusion} proposed to modify the kernel \eqref{eqn:gausskernel} { with a renormalization step} equivalent to replacing~\eqref{eqn:gausskernel} by the right-normalized kernel 
\begin{equation}
\label{eqn:kerna}
e^{-\frac{||x - x_j||^2}{2\epsilon }}\rho_{\epsilon}^{-\alpha}(x_j),
\end{equation}
where $\rho_{\epsilon}$ is the density estimate obtained using \eqref{eqn:kde}. According to \eqref{eqn:gaussian_expansion}, $\rho_{\epsilon}$ has the expansion $\rho_{\epsilon}(x) = \rho(x) + O(\epsilon)$. Then a calculation similar to \eqref{eqn:Markov1}-\eqref{eqn:gensimple} results in the following family of generators
\begin{equation}
\label{eqn:olan}
\mathcal{L}_{\epsilon,\alpha}:=\lim_{\epsilon\rightarrow 0}\frac{\mathcal{P}_{\epsilon,\alpha}f - f}{\epsilon} = \Delta f + \nabla f\cdot\nabla\left[\log\rho^{2(1-\alpha)}\right],
\end{equation}
where $\mathcal{P}_{\epsilon,\alpha}$ is the Markov operator 
for the kernel \eqref{eqn:kerna}. It is easy to check that $\alpha=0.5$ yields the generator for the overdamped Langevin dynamics. Also note that if $\alpha = 1$, the action of Laplace's operator on $f$ is approximated. 

By introducing extra renormalizations to the kernel matrix, one can obtain approximations to more complicated diffusion processes and for data sampled from an arbitrary density.~\cite{coifman2006diffusion,banisch2020,trstanova2020local,evans2021computing}

\subsubsection{Target Measure Diffusion Map}
In order to quantify rare transitions between metastable regions using the TPT framework, one needs to approximate the generator of the process within the reactive channels and solve the committor boundary value problem \eqref{eqn:comm_pde}. Hence, it is necessary to collect data there. This is typically done using \emph{enhanced sampling}: sampling an auxiliary dynamics where rare events are no longer rare. Diffusion maps is a powerful but sampling-dependent method, and enhanced sampling techniques are needed to fuse the algorithm with meaningful study of rare events in molecular simulations. One natural solution is to choose a distribution that is easier to sample, e.g from a higher temperature system~\cite{so2000temperature,abrams2010large} or from a `flat' energy landscape,~\cite{wang2000} and then reweight to a \emph{target measure} $\mu(x).$ This reweighting is the idea of \emph{importance sampling} for Monte Carlo integration, and also the foundation of umbrella sampling,~\cite{torrie1977nonphysical} one of the oldest enhanced sampling methods.

The target measure diffusion map algorithm~\cite{banisch2020} ({\tt tm-dmap}) modifies the right normalization of the kernel to incorporate a user-provided target measure $\mu(x).$ The target measure does not need to be normalized and it can be given in the form of a vector $\{\mu(x_i)\}_{i=1}^N$ defined on the dataset. The dataset itself has a sampling density $\rho(x)$ (not necessarily known) which {\tt tm-dmap} reweights to replace with $\mu(x).$ 
The generator approximated by {\tt tm-dmap} corresponds to the overdamped Langevin dynamics 
\begin{equation}
\label{eqn:overdamped}
dx = -\nabla  F(x)dt +\sqrt{2\beta^{-1}}dw
\end{equation}
with stationary distribution $\mu(x)= \exp(-\beta F(x)),$ and is independent of the sampling density.
For use in metastable systems in molecular dynamics, one can generate data from an enhanced sampling procedure, run diffusion maps on the data, and reweight the results to a provided target measure such as a specified Gibbs density. 


The {\tt tm-dmap} makes use of the kernel density estimate $\rho_{\epsilon}(x)$ using~\eqref{eqn:kde}.
Then \emph{right-normalized} kernel is defined for a pair of data points $x,x'$ as
\begin{equation}
\label{eqn:rnkernel}
k_{\epsilon,\mu}^{\tt tm-dmap}(x,x'):=\frac{ e^{-\frac{\|x-x'\|^2}{2\epsilon}}\mu^{1/2}(x')}{\rho_{\epsilon}(x')}.
\end{equation}
In the denominator, the power $1$ of the sampling density estimate leads to its complete cancellation in the order $\epsilon$ term in a calculation similar to \eqref{eqn:Markov1}--\eqref{eqn:Markov2}. In the numerator, the power $1/2$ of the target measure $\mu$ results in the order $\epsilon$ term being $\Delta f + \nabla f\cdot \log \mu$ which is $\beta/2$ times the generator for \eqref{eqn:overdamped}.




\section{Proposed Methodology}\label{sec:methodology}

\subsection{Target Measure Mahalanobis Diffusion Map}\label{sec:tm-mmap}

An important limitation of 
the target measure diffusion map ({\tt tm-dmap})~\cite{banisch2020,trstanova2020local}
is that it only approximates generators that
are relevant for gradient flows \eqref{eqn:overdamped}. 
The {\tt tm-dmap} algorithm utilizes 
the user-input target measure, but not dynamical properties of the data. Note that the invariant measure for the dynamics in collective variables \eqref{eqn:cvsde} with position-dependent diffusion tensor is the same as for the overdamped Langevin dynamics \eqref{eqn:overdamped}: $\exp(-\beta F)$.

Position-dependent diffusion
is essential in dynamical models for collective variables,  
tracing back to Kramer's model for diffusive barrier crossing.~\cite{kramers1940brownian}
This diffusion naturally arises from the coupling of collective variables to the solvent~\cite{straub1987calculation,johnson2012characterization} and heavily influences the reaction pathway and reaction coordinate in problems such as protein folding~\cite{best2010} and membrane permeation.~\cite{sicard2021position}

In our recent work,~\cite{evans2021computing} we proved that the Mahalanobis diffusion map algorithm ({\tt mmap}) approximates the generator for the overdamped Langevin dynamics in collective variables \eqref{eqn:mgen} with $O(\epsilon)$ accuracy and applied {\tt mmap} to collective variable data  of MD simulations of Lennard-Jones-7 in 2D and of the transition between metastable states C7eq and C5 of alanine dipeptide in vacuum. Importantly, we were not able to apply {\tt mmap} to the transition C7eq--C7ax in alanine dipeptide because of the lack of data in the transition region sampled from the invariant density required for {\tt mmap}.  

The {\tt mmap} algorithm uses the Mahalanobis kernel introduced by Singer and Coifman~\cite{singer2008}
\begin{equation}
\label{eqn:mahal_kernel}
 k_{\epsilon}^{\tt mmap}(x,x') = e^{-{\frac{(x-x')^{\top}[M^{-1}(x) + M^{-1}(x') ](x-x')}{4\epsilon}}},
\end{equation}
where each $M(x)$ is a user-input $d \times d$ symmetric positive-definite matrix considered as the diffusion matrix for data point $x.$ 
The difference between each component of $x$ and $x'$ is normalized by the corresponding variance and reflects the difficulty to deviate along each direction. Therefore, the Mahalanobis kernel~\eqref{eqn:mahal_kernel} is a decaying exponential of Mahalanobis distance squared, and is designed to account for anisotropy of the diffusion process the data is coming from. 

{ An important difference between the cases considered in Ref.~\onlinecite{singer2008} and our work~\cite{evans2021computing} is that the diffusion matrix $M(x)$ in Ref.~\onlinecite{singer2008} is of the form $JJ^\top(x)$ where $J$ is the Jacobian matrix for some diffeomorphism, while we have omitted this requirement. Indeed, the diffusion tensors computed using restrained dynamics and averaging~\cite{2006string} are not of the form $JJ^\top(x)$.~\cite{evans2021computing}
}



In this work, we extend the results of Ref.~\onlinecite{evans2021computing} to target measure diffusion map resulting in the \emph{target measure Mahalanobis diffusion map} algorithm, abbreviated as {\tt tm-mmap}. The key modification in {\tt tm-mmap} in comparison with {\tt tm-dmap} is that the right-normalized kernel function \eqref{eqn:rnkernel} is changed to 
\begin{equation}
\label{eqn:rnmkernel}
 k_{\epsilon, \mu}^{\tt tm-mmap}(x,x') = k_{\epsilon}^{\tt mmap}(x,x')\frac{\mu(x')^{1/2}| M(x')|^{-1/4}}{\rho_{\epsilon}(x')},
\end{equation}
where $k_{\epsilon}$ is the Mahalanobis kernel \eqref{eqn:mahal_kernel}, $|M(x')|$ denotes the determinant of $M(x)$ { and $\rho_{\epsilon}(x)$ is defined as~\eqref{eqn:mmap_density_original} with $c_{\epsilon}(x):= (2\pi \epsilon)^{d/2}|M|^{1/2}(x).$}
The {\tt tm-mmap} algorithm written out in the panel Algorithm \ref{alg:tmap} follows the steps of {\tt tm-dmap}.~\cite{banisch2020} 

\begin{algorithm}
  \SetAlgoNoLine
  \DontPrintSemicolon
  \KwIn{data $X = \{x_i\}_{i=1}^N$, diffusion matrices $\{M(x_i)\}_{i=1}^N$ bandwidth $\epsilon$, target measure $\mu(x_i)$}
  \KwOut{Generator matrix $L_{\epsilon,\mu}$}
  \nonl \custom{\textnormal{Construct kernel from~\eqref{eqn:mahal_kernel}, estimate sampling density}}{
  $\displaystyle [K_{\epsilon}]_{i,j}  = k_{\epsilon}^{\tt mmap}(x_i, x_j), i,j=1,\ldots,N $ \;
  $\displaystyle c_i = (2\pi\epsilon)^{d/2} |M_i|^{1/2},$ $i=1,\ldots, N$\;
  $\displaystyle [p_{\epsilon}]_i = \frac{1}{N c_i}\sum_j [K_{\epsilon}]_{ij}, i=1,\ldots, N$
  }
  \nonl\custom{\textnormal{Right normalize the kernel}}{
  $\displaystyle[K_{\epsilon,\mu}]_{ij} := \frac{[K_{\epsilon}]_{ij} (\mu_j |M_j|^{-1/2})^{1/2}}{[p_{\epsilon}]_j}, \ i,j = 1,\ldots, N$ }
   \nonl\custom{\textnormal{Left normalize the kernel}}{
  $\displaystyle [P_{\epsilon, \mu}]_{ij}:= \frac{[K_{\epsilon,\mu}]_{ij}}{ \sum_{\ell}[K_{\epsilon,\mu}]_{i\ell}}, \ i,j=1,\ldots,N  $}
  \nonl\custom{\textnormal{Construct generator}}
  {$\displaystyle [L_{\epsilon,\mu}]_{ij} = \frac{[P_{\epsilon,\mu}]_{ij} - \delta_{ij}}{\epsilon}, i,j=1,\ldots, N$\;}
  \caption{Target Measure Mahalanobis Diffusion Map ({\tt tm-mmap}) }
  \label{alg:tmap}
\end{algorithm}

Suppose that we have a sampling procedure that allows us to generate a dataset $\{x_i\}_{i=1}^N$ of any size $N$ distributed according to an arbitrary measure {$\rho(x)$ with the property that the target measure $\mu(x)$ is absolutely continuous with respect to $\rho$. This means $\rho(A) = 0$ implies that $\mu(A) = 0$ for any measurable set $A$. 
The main theoretical result of this work is that \emph{Algorithm \ref{alg:tmap} constructs an approximation to the generator $\mathcal{L}$ \eqref{eqn:mgen} for the overdamped Langevin dynamics in collective variables \eqref{eqn:cvsde}}. More precisely, suppose we are given
\begin{itemize}
    \item a smooth target measure $\mu(x) = e^{-\beta F(x)}$, 
    \item a smooth symmetric positive definite diffusion matrix-function $M(x)$ with determinant bounded away from zero and entries of $M^{-1}(x)$ being smooth and bounded functions, and
    \item a smooth and bounded function $f(x)$.
\end{itemize}
Then the following limit takes place for all $x_i$, $1\le i\le N$, that have enough neighbors in their neighborhoods of radius $\sim3\sqrt{\epsilon}$:
\begin{equation}
\label{eqn:limN}
\lim_{\epsilon\rightarrow 0}\lim_{N\rightarrow\infty} [L_{\epsilon,\mu}[f]]_i = \lim_{\epsilon\rightarrow 0}[\mathcal{L}_{\epsilon,\mu}f](x_i) = \frac{\beta}{2}[\mathcal{L}f](x_i),
\end{equation}
where $[f] = [f(x_j)]_{j=1}^N$,
the matrix operator $L_{\epsilon, \mu}$ is constructed according to Algorithm~\ref{alg:tmap}, the integral operator $\mathcal{L}_{\epsilon,\mu}$ is the continuous counterpart of $L_{\epsilon, \mu}$, and $\mathcal{L}$ is the generator \eqref{eqn:mgen}.  The limit 
$$
\lim_{N\rightarrow\infty} [L_{\epsilon,\mu}[f]]_i = [\mathcal{L}_{\epsilon,\mu}f](x_i)
$$ 
follows from the law of large numbers. Indeed, 
\begin{align*}
[L_{\epsilon,\mu}[f]]_i  & = \frac{[P_{\epsilon,\mu}[f]]_{i} - [f]_i}{\epsilon} \\
& = \frac{1}{\epsilon}\left(\frac{\sum_{j=1}^N [K_{\epsilon,\mu}]_{ij}[f]_j}{\sum_{j=1}^N [K_{\epsilon,\mu}]_{ij}} - [f]_i\right) \\
& \rightarrow 
\frac{1}{\epsilon}\left(\frac{\int_{\Omega}k_{\epsilon, \mu}^{\tt tm-mmap}(x_i,x')\rho(x')f(x')dx'}{\int_{\Omega}k_{\epsilon, \mu}^{\tt tm-mmap}(x_i,x')\rho(x')dx'} - f(x_i)\right)\\
&=: [\mathcal{L}_{\epsilon,\mu}f](x_i).
\end{align*}
The limit $\epsilon\rightarrow 0$ is established as a result of a tedious calculation consisting of two stages. First, a Taylor expansion in $\epsilon$ for the integrals of the form
$$
\int_{\Omega}k_{\epsilon}^{\tt mmap}(x,x')f(x')dx'
$$
is computed. This is done in Ref.~\onlinecite{evans2021computing} (see Appendices C and D) and the result is stated here in Lemma \ref{thm:lem2} in Appendix \ref{sec:appendixA}. Second, a calculation similar to \eqref{eqn:Markov1}--\eqref{eqn:gensimple} is conducted -- see Theorem \ref{thm:main-theorem} in Appendix \ref{sec:appendixA}.}

\subsection{Transition Path Theory in collective variables with diffusion maps}
\label{sec:TPTCV}
To obtain the committor, we solve the matrix equation
\begin{align}
  [L_{\epsilon,\mu}q]_i & = 0, \quad  x_i \in \Omega \backslash (A \cup B),
  \label{eqn:comm_discrete}\\
  [q]_i  & = 0,  \quad  x_i \in A, \nonumber                 \\
  [q]_i  & = 1,  \quad x_i \in B \nonumber.
\end{align}
{The matrix in~\eqref{eqn:comm_discrete} is sparsified -- see Section \ref{sec:DMimp} -- allowing for the use of sparse linear algebra solvers. 
}   

Since $L_{\epsilon, \mu}$ approximates the generator $\mathcal{L}$ for the overdamped Langevin dynamics in collective variables from \eqref{eqn:cvsde}, the solution to \eqref{eqn:comm_discrete} converges to the solution to the corresponding committor problem \eqref{eqn:comm_pde} where $\mathcal{L}$ is given by \eqref{eqn:mgen}.
Once the committor is found, one can calculate 
the probability density of reactive trajectories, the reactive current on the data, and the transition rate.

To compute the reactive current, we need an estimate for the target density $\pi(x) := Z^{-1}\mu(x),$ with $Z = \int_{\R^d} \mu(x) dx.$ If we don't have access to $Z,$ we can estimate this with the kernel density estimate $p_{\epsilon}$ from~\eqref{eqn:kde}.
Namely, we estimate $Z$ via Monte Carlo integration with kernel reweighting as
\begin{equation}
\label{eqn:Ze}
Z_{\epsilon} := \frac{1}{N}\sum\limits_{i=1}^{N} 
\frac{\mu(x_i)}{[p_{\epsilon}]_i}
\end{equation}
as for sufficiently large $N$ and small $\epsilon$ we have
\[Z_{\epsilon}  \approx \int_{\R^d} \frac{\mu(x)}{\rho(x)} \rho(x)dx = Z.\]

{ Then the discrete reactive current and the transition rate can be calculated directly on the point cloud without meshing it, as described in Appendix E of Ref.~\onlinecite{evans2021computing}. The formulas for the reactive current \eqref{eqn:cvrcur} and the transition rate \eqref{eqn:nuAB1} include the term $M\nabla q$. The key observation facilitating this calculation~\cite{pavliotis2014stochastic} is that for any two functions $g_1$ and $g_2$  and the generator \eqref{eqn:mgen_divergence} for SDE \eqref{eqn:cvsde} we have:
\begin{equation}\label{eqn:product_rule}
\mathcal{L}_{\epsilon,\mu}(g_1g_2) - g_1\mathcal{L}_{\epsilon,\mu}g_2 - g_2\mathcal{L}_{\epsilon,\mu}g_1 = \frac{2}{\beta}\nabla g_1 ^{\top} M \nabla g_2.
\end{equation}
On the other hand, using Monte Carlo integration and simple algebra, recalling that $\mathcal{L}_{\epsilon,\mu}$ approximates $\tfrac{\beta}{2}\mathcal{L}$, we obtain:
\begin{equation}
\label{eqn:aux1}
\nabla g_1 ^{\top} M \nabla g_2 \approx \sum_{j=1}^N[L_{\epsilon,\mu}]_{ij}([g_1]_i-[g_1]_j)([g_2]_i-[g_2]_j).
\end{equation}
}
The discrete reactive current $\hat{\mathcal{J}_{\epsilon}} \in \R^{d \times N}$ is obtained using \eqref{eqn:cvrcur}, \eqref{eqn:Ze}, and \eqref{eqn:aux1} with $[g_1] = [x]_l$, the $l$th entry of $x$, and $[g_2] = [q]$: 
\begin{flalign}
    \label{eqn:discrete_current}
    [\hat{\mathcal{J}_{\epsilon}}]_{i,l} :=
    \frac{ \mu(x_i)}{\beta Z_{\epsilon} }\sum_{j=1}^N [L_{\epsilon}]_{ij}([q_{\epsilon}]_i - [q_{\epsilon}]_j)(x_{i,l} -
    x_{j,l}).
\end{flalign}
Here,  $[\hat{\mathcal{J}_{\epsilon}}]_{i,l}$ denotes the $l$th component of the current at $x_i$, and $x_{i,l}$ is the $l$th entry of $x_i$, $1\le l\le d$, $1\le i\le N$.

The reaction rate $\hat{\nu}_{AB}$ is estimated using \eqref{eqn:nuAB1}, \eqref{eqn:Ze}, and \eqref{eqn:aux1} with $[g_1] = [g_2] = [q]$:
\begin{equation}
    \label{eqn:rate_gamma_reweight}
\hat{\nu}_{AB} = \frac{1}{|I_{AB}|} \sum\limits_{i \in I_{AB}} \sum_{j=1}^N
[L_{\epsilon}]_{ij}([q_{\epsilon}]_i - [q_{\epsilon}]_j)^2 \frac{ \mu(x_i)}{Z_{\epsilon}[p_{\epsilon}]_i}
\end{equation}
where $I_{AB} = \{i: x_i \in \Omega
\backslash (A \cup B)\}$. 
\subsection{Choosing $\epsilon$ }
\label{sec:heuristic}
In practice, the limit $\epsilon \to 0$ cannot be taken for a finite dataset. Instead, one generally tries to choose $\epsilon$ as small as possible { without making the corresponding generator matrix $L_{\epsilon,\mu}$ reducible.}
We note that many heuristics exist for choosing the scaling parameter $\epsilon$ in diffusion maps, relating back to bandwidth selection in kernel density estimation.~\cite{lindenbaum2020gaussian} For this work, we choose the method of Berry, Harlim and Giannakis~\cite{berry2015nonparametric,berry2016variable,giannakis2019data,davis2021graph} which we refer to as the  \emph{Ksum test}.

The idea for the heuristic is to find the range of $\epsilon$ where the asymptotic results of diffusion maps hold true for the given dataset. We find this range by analyzing the double sum 
$$
S(\epsilon):={\frac{1}{N^2}}\sum_{i=1}^N\sum_{j=1}^N [K_{\epsilon}]_{ij}
$$ 
 over a range of $\epsilon$ values. Here, $[K_{\epsilon}]_{ij} = k_{\epsilon}(x_i,x_j)$ where $k_{\epsilon}$ is the Mahalanobis kernel \eqref{eqn:mahal_kernel}.
{For large $N$, the intermediate asymptotic for $S(\epsilon)$ is ~\cite{coifman2008graph,berry2016variable}
$$
S(\epsilon)\approx\int_{\Omega^2}k_{\epsilon}(x,x')dxdx' \approx C\epsilon^{d/2}
$$
where $C$ is a constant. Note that $C$ depends on the distance function in the kernel and the volume of $\Omega$ that is assumed to be finite, but is independent of  $\epsilon$. Hence, 
\begin{equation}
\label{eqn:logS}
\log S(\epsilon) \approx \frac{d}{2}\log \epsilon + \log C.
\end{equation}

Therefore, if we plot  $\log S(\epsilon)$ against $\log \epsilon$ we should see a linear region of slope approximately $\tfrac{d}{2}$, where $d$ is the dimension of the dataset. { This region demarcates the range of suitable values of $\epsilon$.~\cite{berry2015nonparametric,berry2016variable,giannakis2019data,davis2021graph} 
On the other hand,  if $\epsilon$ is large,  $[K_{\epsilon}]_{ij} \approx 1$ for all $i,j,$ and hence $S(\epsilon) \to 1$ as $\epsilon\rightarrow \infty$.} For small $\epsilon,$
$[K_{\epsilon}]_{ij} \approx 0$ for all $i,j$, $i \neq j$, and $[K_\epsilon]_{ii} = 1$. Therefore, $S(\epsilon) \to N^{-1}$ as $\epsilon \to 0.$
Hence, the slope of the graph of $\log S(\epsilon)$ versus $\log \epsilon$ should tend to zero as $\epsilon\rightarrow 0$ and as $\epsilon\rightarrow\infty$. Therefore, the Ksum test suggests to choose the value of $\epsilon$ corresponding to the maximal slope of this graph.}

For practical calculation, it is useful to note that
{\begin{equation}
  \label{eqn:logS_formula}
  \frac{\partial \log S(\epsilon)}{\partial \log \epsilon} = \frac{\frac{\partial S(\epsilon)}{\partial \epsilon}\frac{d \epsilon}{ d\log\epsilon}}{S(\epsilon)}  
  = -\frac{\sum_i\sum_j [K_{\epsilon}]_{ij} \log [K_{\epsilon}]_{ij}}{\sum_i\sum_j [K_{\epsilon}]_{ij}}.
\end{equation}
}
The procedure for choosing a good bandwidth $\epsilon^{\ast}$ is outlined in Algorithm~\ref{alg:kdoublesum} in Appendix~\ref{sec:algorithms}.


\subsection{Diffusion map implementation}
\label{sec:DMimp}
The approximation error for diffusion maps scales as $O(N^{-1/2})$ where $N$ is the number of data points. On the other hand, the kernel matrix is dense as the exponential function in the kernel function is always positive, though exponentially decaying. Therefore, it is important to  utilize a sparse approximation $\tilde{K}_{\epsilon}$ for the kernel matrix. Typically practitioners will use a $k$-nearest neighbors (KNN) approach or a radius-nearest neighbors (RNN) approach. We prefer the RNN-based sparse approximation as it gives results consistent with our expectations even if the choice of the bandwidth is poor and it does not require symmetrization of $\tilde{K}_{\epsilon}$.
The radius nearest neighbors approach uses
\begin{equation}
    \tilde{k}_{\epsilon}(x, x') = \begin{cases}
	k_{\epsilon}(x,x'), & d(x,x') \leq r \\
    0, & {\rm otherwise} \\
	\end{cases},
\end{equation}
where $d(x,x')$ is the distance used in the kernel and $r$ is typically chosen to be $3\sqrt{\epsilon}$.
 
To implement {\tt tm-mmap} where data points $x, x'$ are on a torus (such as for dihedral angles), we compute the Mahalanobis kernel~\eqref{eqn:mahal_kernel} { accounting for the periodic boundary conditions} in computing the distance vector $x - x'$.

\subsection{Metadynamics}
\label{sec:metad}
For enhanced sampling we utilize the well-tempered metadynamics algorithm (WTMETAD).~\cite{barducci2008} WTMETAD is an adaptive enhanced sampling algorithm which updates the potential of the simulation as the trajectory progresses. At the $n$th biasing step, the existing bias at the current location $x_n$ is contributed to the bias term
\begin{equation}
  \label{eqn:metad_update}
  \hspace{-0.05cm}V_{n}(x) = V_{n-1}(x) + \Big[he^{-(x - x_n)^{\top}\Sigma^{-1}(x - x_n)} \Big] \Big[e^{-\frac{\beta}{\gamma - 1}V_{n-1}(x_{n})}\Big]
\end{equation}
for the $n$th biasing step. The bias function parameters are the height of the biasing kernel $h,$ the covariance matrix $\Sigma$, and the biasing factor $\gamma.$ Usually, the covariance is chosen to be a diagonal matrix $\Sigma = \text{diag}(\sigma_1,\ldots,\sigma_d)$ where $\sigma_j$, $1\le j\le d$ are based on the unbiased CV fluctuations. The \emph{well-tempered} variant of metadynamics is distinguished by the factor 
$\exp\left(-\tfrac{\beta}{\gamma-1}V_{n-1}(x_{n})\right)$ which scales the Gaussian biases. In the limit $\gamma \to 1,$ the well-tempered term tends to $0$ and there is no biasing, while for $\gamma \to \infty$ we have the well-tempered term tending to $1$ and recover the usual metadynamics algorithm.
The user also chooses a timestep $t_{bias} = n_{bias} \Delta t,$ where $\Delta t$ is the timestep of the simulation.
As the simulation runs, the height of the deposited Gaussian bumps decreases. As a result, the total biasing potential becomes less rugged as the metadynamics run progresses. Moreover, it converges to $-(1 - \frac{1}{\gamma})F(x),$ particularly the negative of the free energy as $\gamma \to \infty$. Hence, in the long run, metadynamics flattens out the free energy and with large $\gamma$ will fill its level sets uniformly with samples.
{We utilize well-tempered metadynamics because of its convergence guarantees~\cite{fort2018convergence,valsson2016,dama2014well}} and its improved reweighting over the standard metadynamics.~\cite{tiwary2015time}

\subsection{Obtaining the free energy and diffusion matrix in high dimensions}
\label{sec:FE&M}
Besides the dataset, {\tt tm-mmap} requires the target measure and the diffusion matrix evaluated at the data points. The evaluation of the diffusion matrix is computationally demanding but straightforward using local restrained simulations.~\cite{2006string} The evaluation of the target measure if it is not known a priori is less straightforward. We propose a method for estimating  the target measure $\mu = \exp(-\beta F(x))$ in high dimensions that works with any standard enhanced sampling algorithm with known biasing potential. This method is inspired by Ref.~\onlinecite{tiwary2015time}.

Suppose that an enhanced sampling algorithm samples from the Gibbs density
$$
\rho(x) \propto e^{-\beta(F(x) + U(x))},
$$
where $U$ is a \emph{known} bias potential, while
$F(x)$ is the \emph{unknown} free energy.
The desired target measure is $\mu(x) =\exp(-\beta F(x)).$ 
We approximate the sampling density $\rho(x)$ by $\rho_{\epsilon}(x)$ given by \eqref{eqn:kde} and obtain the following estimate for the target measure:
\begin{equation}
\label{eqn:mu_estimate}
\mu(x) \approx \rho_{\epsilon}(x)\exp(\beta U(x)).
\end{equation}

\subsection{Preparing the dataset}
\label{sec:prep_dataset}
A remarkable property of {\tt tm-mmap} is that it allows for the use of input data distributed according to an arbitrary measure $\rho$ provided the target measure $\mu$ is absolutely continuous with respect to $\rho$. In practice, this means that the data points should cover all regions where the target measure $\mu$ is above a certain threshold which should be strictly smaller than the value of the target measure at the transition state. Therefore, {\tt tm-mmap} enables us not only to generate data points using any enhanced sampling algorithm but also post-process them as we see fit. In this work, we used two enhanced sampling algorithms to generate data: temperature acceleration and well-tempered metadynamics (see Section \ref{sec:metad}). We generated long trajectories using these enhanced sampling algorithms and then post-processed them to leave roughly $N = 10^4$ points as follows:
\begin{itemize}
    \item Trajectory data generated using temperature acceleration were subsampled uniformly in time.
    \item Trajectory data generated by well-tempered metadynamics were subsampled 
    \begin{itemize}
        \item uniform in time
        \item quasi-uniform in space resulting in the so-called \emph{delta-net}.~\cite{crosskey2017atlas}       
    \end{itemize}
\end{itemize}
We compare the performance of {\tt tm-mmap} on these three types of datasets and show that the the delta-net data leads to the most accurate and the most robust results. The construction of the delta-net is detailed in Algorithm \ref{alg:deltanet} in Appendix~\ref{sec:algorithms}.

\section{Numerical Tests}
\label{sec:examples}
In this section, we test {\tt tm-mmap} on two examples: the Moro-Cardin two-well system with position-dependent diffusion, and on alanine dipeptide with two dihedral angles in vacuum. The results of {\tt tm-mmap} are compared to those computed by means of the finite element method (FEM). We weigh the numerical error according to the probability density of reactive trajectories:
\begin{align}
E& = \sum\limits_{i=1}^{N}|q(x_i) - q_{FEM}(x_i)|w(x_i),\label{eqn:WAE}\\
w(x_i) & =\frac{ \mu(x_i)q(x_i)(1-q(x_i))}
{\sum_j \mu(x_j)q(x_j)(1-q(x_j)) }. \label{eqn:w}
\end{align}
The resulting error will be abbreviated as WAE (Weighted Absolute Error).

{ An application of {\tt tm-mmap} to alanine dipeptide with four dihedral angles will be reported in Section \ref{sec:aladip4D}.}

\subsection{The Moro-Cardin system}
\label{sec:MoroCardin}
\begin{figure}
\centering
    \includegraphics[width = 0.45\textwidth]{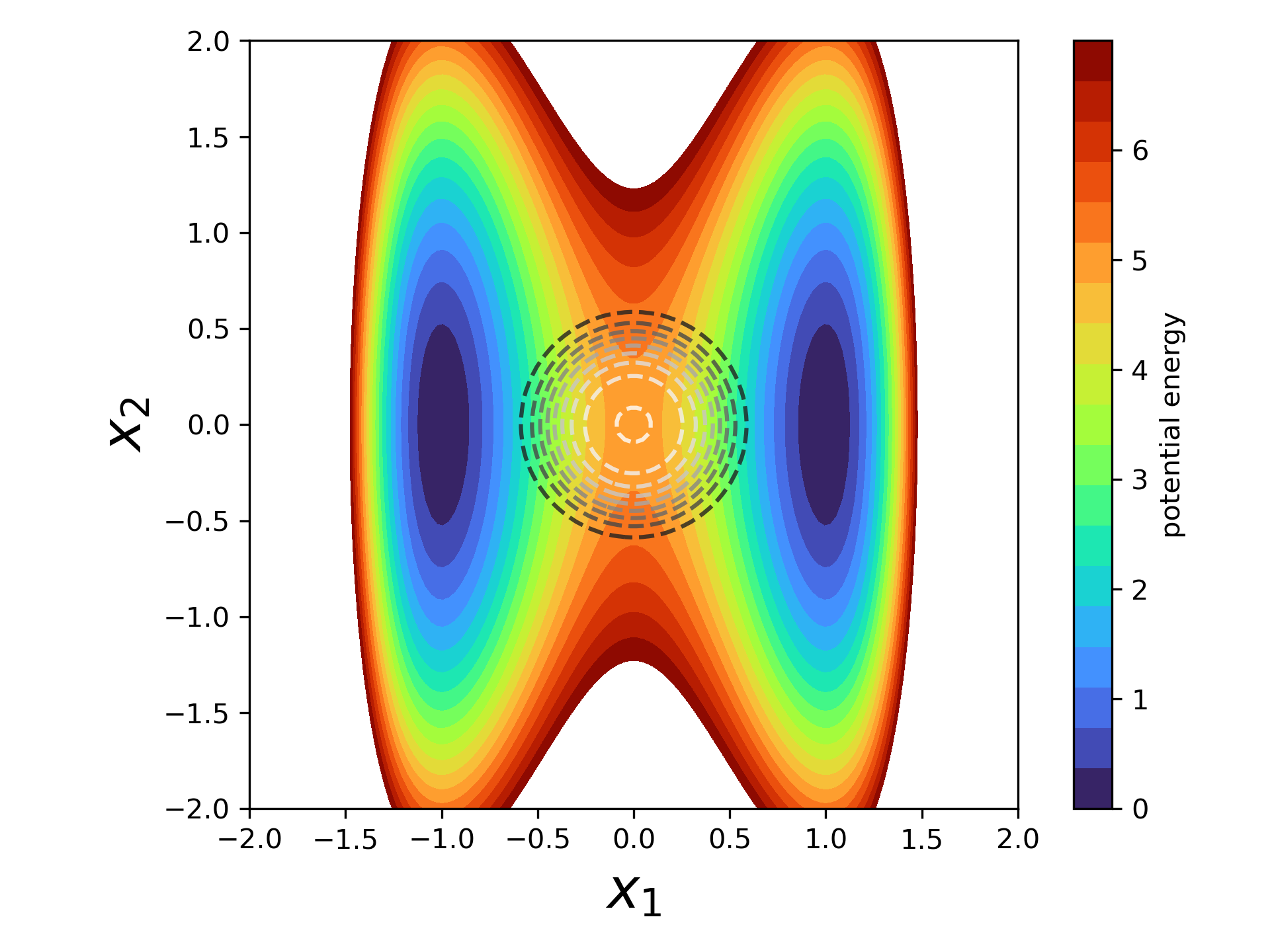}
    \caption{The potential energy function (color coding) and the level sets of the diffusion coefficient (dashed circles of grey shades) for the Moro-Cardin system. The level sets of the diffusion coefficient indicate a low-diffusivity region around the origin.
    }
\label{fig:morocardin_potential_diffusion}
\end{figure}

The Moro-Cardin two-well system with position-dependent diffusion~\cite{moro1998saddle} is a simple and insightful example featuring \emph{saddle avoidance}. The system evolves according to SDE \eqref{eqn:cvsde} with the potential function  $F(x)\equiv V(x)$ and diffusion matrix $M(x)$ given by
\begin{align} 
  V(x) & = 5(x_1^2-1)^2 + 10\alpha x_2^2,  \label{eqn:morocardin_potential}\\
  M(x) & = (1 + 8e^{-||x||^2/2\sigma^2})^{-1}I_{2\times2},\label{eqn:morocardin_M}
\end{align}
where $\sigma = 0.2$ and $\alpha = \text{arctan}(7\pi/9)$. The inverse temperature-like parameter $\beta$ in SDE \eqref{eqn:cvsde} is chosen to be 1.  The diffusion matrix $M(x)$ is close to the identity matrix everywhere except for a ball around the origin of radius $2\sigma = 0.4$, while it decays to $(1/9)I_{2\times2}$ at the origin. The function $V(x)$ has two minima at $(-1,0)$ and $(1,0)$, and a saddle at the origin. The sets $A$ and $B$ are chosen to be balls centered at the minima of radius 0.2. Level sets of the potential function and the diffusion coefficient are depicted in Figure \ref{fig:morocardin_potential_diffusion}.

\begin{figure}
\centering
    \includegraphics[width = 0.45\textwidth]{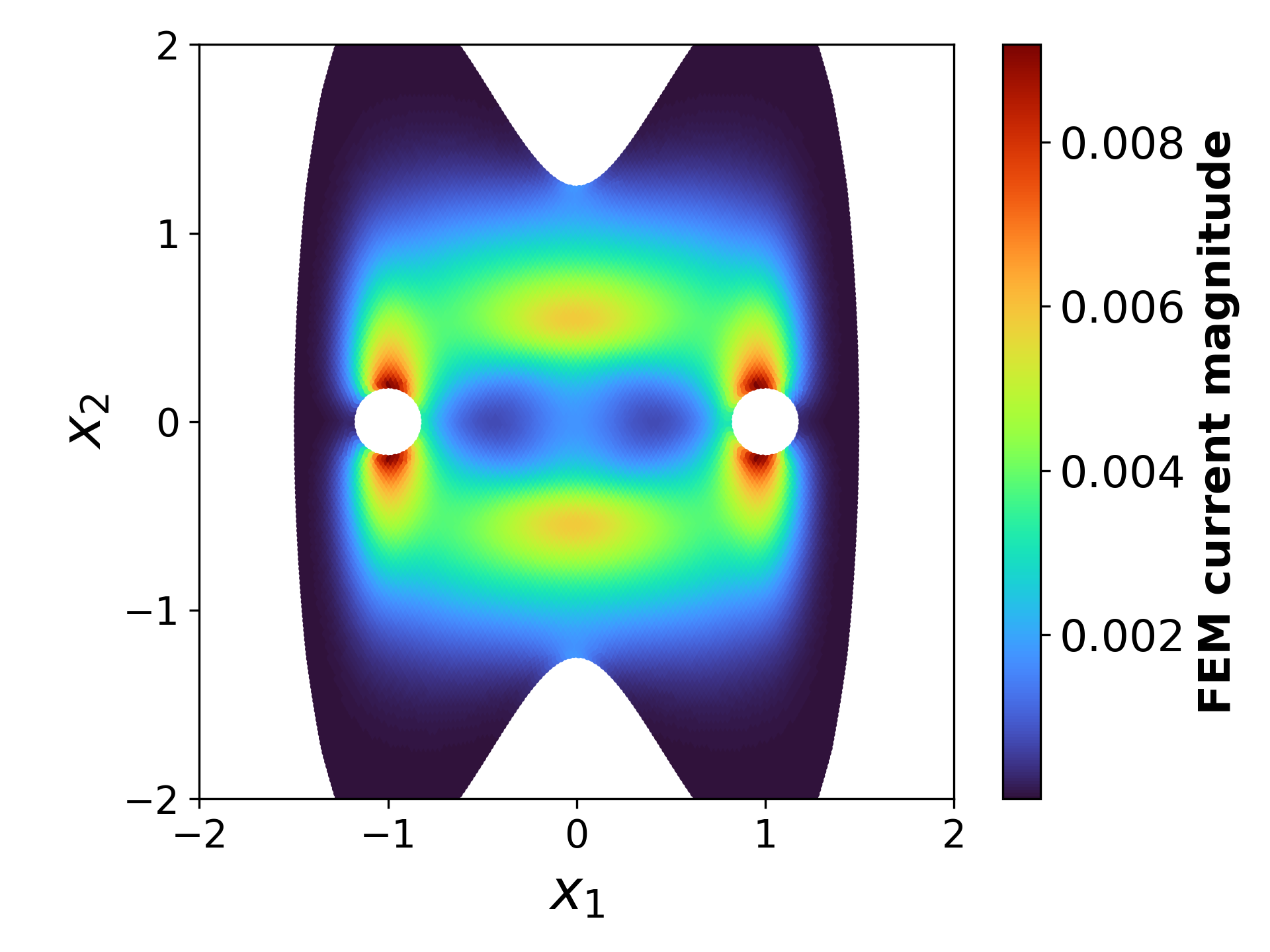}
    \caption{The reactive current computed for the Moro-Cardin system (Section \ref{sec:MoroCardin}) using the finite element method.
    }
\label{fig:MCfem}
\end{figure}

The dynamics of the Moro-Cardin system { manifest} saddle avoidance (see Figure \ref{fig:MCfem}). Those reactive trajectories that pass near the saddle are slowed down by low diffusivity so much that they become trapped in this region for a long time. { Consequently}, the reactive trajectories that climb over higher barriers and avoid the low-diffusivity region surrounding the origin will {surpass those that} go over a lower barrier but get trapped in the low-diffusivity region. As a result, the reactive flux splits into two reaction channels bending around the origin from above and from below.

We created three datasets as described in Section \ref{sec:prep_dataset}. The first dataset, denoted by {\tt hightemp}, was generated by running a long trajectory using the Euler-Maruyama method with time step $dt = 10^{-4}$ for $10^6$ steps at a high temperature $\beta^{-1} = 3$, and then subsampling the trajectory at equispaced time intervals.  The second and third datasets were generated using WTMETAD with parameters $h=0.35,$ $\sigma_1=\sigma_2=0.1$, $\gamma=5.0$ and deposition rate of every 500 timesteps (see Section \ref{sec:metad}). Then the datasets {\tt metad} and {\tt deltanet} were obtained, respectively, by subsampling WTMETAD trajectory data uniformly in time and quasi-uniformly in space using Algorithm \ref{alg:deltanet} with $\delta = 0.017$. All datasets contain a total of $N = 10^4$ data points.

For each of these datasets, we computed the matrix operator $L_{\epsilon,\mu}$ using Algorithm \ref{alg:tmap} and then computed the committor, the reactive current, and the transition rate as described in Section \ref{sec:TPTCV}. 

For validation, we also computed the committor, the reactive current displayed in Figure \ref{fig:MCfem}, and the transition rate using the finite element method (FEM) . The points of the finite element mesh were used as the test points $x_i$. The error metric for the committor was the weighted absolute error (WAE) defined in \eqref{eqn:WAE}--\eqref{eqn:w}. 

\begin{figure*}
\centering
\includegraphics[width=\textwidth]{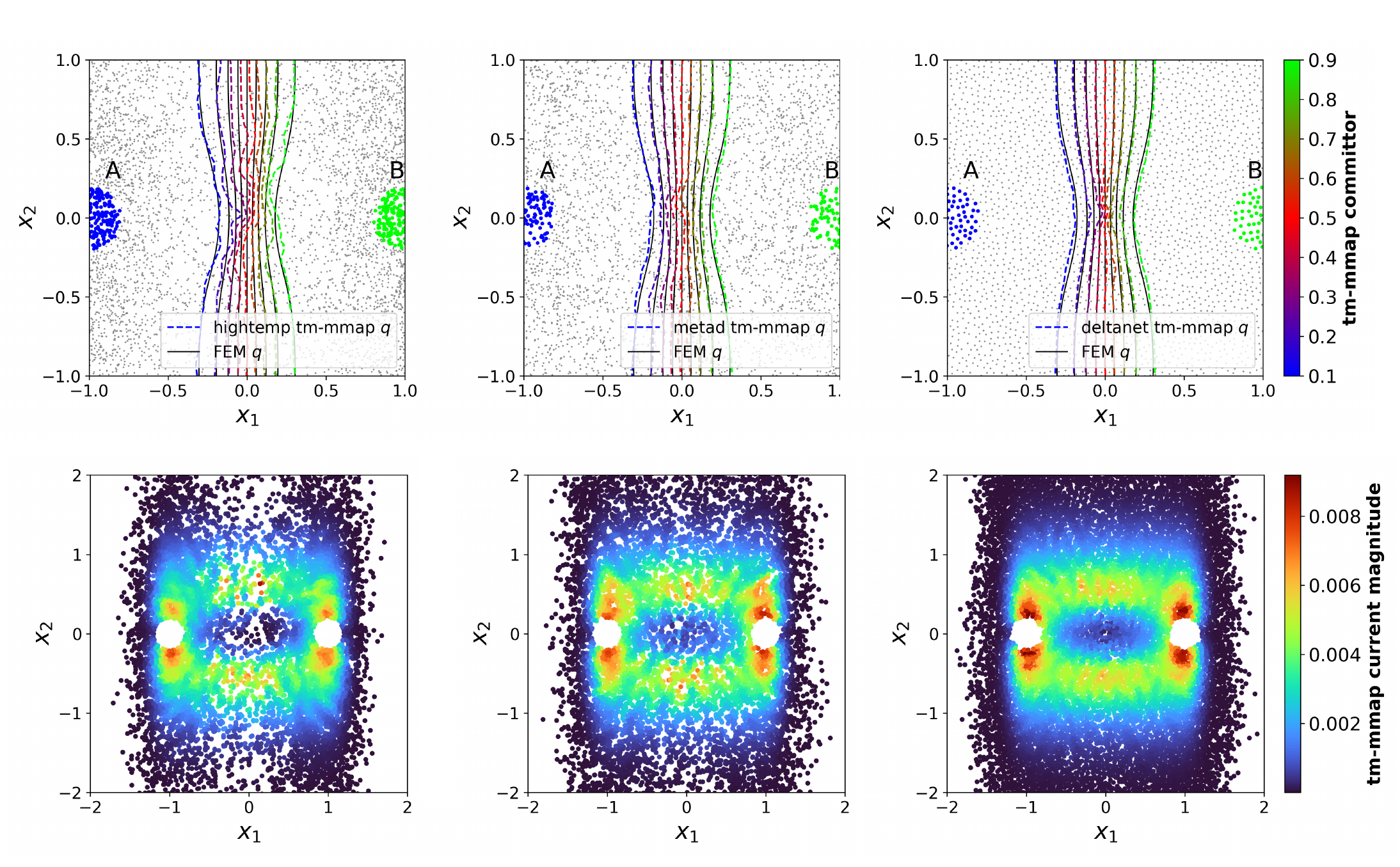}
  \caption{An application of {\tt tm--mmap} to the Moro-Cardin system (Section \ref{sec:MoroCardin}). Committor level sets (top row) and reactive current magnitudes (bottom row) computed by {\tt tm-mmap} using three datasets as input. Left column: dataset {\tt hightemp}, temperature accelerated molecular dynamics data for $\beta^{-1} = 3$ subsampled uniformly in time. Middle column: dataset {\tt metad}, WTMETAD trajectory data subsampled uniformly in time. Right column: dataset {\tt deltanet},  WTMETAD trajectory data subsampled quasi-uniformly in space using Algorithm \ref{alg:deltanet}. The level sets of the committor computed using {\tt tm-mmap} (dashed colored curves) are superimposed with the level sets of the committor computed using the finite element method (solid black curves).}
  \label{fig:morocardin_validation}
\end{figure*}

\begin{figure*}
\centering
    \includegraphics[width=\textwidth]{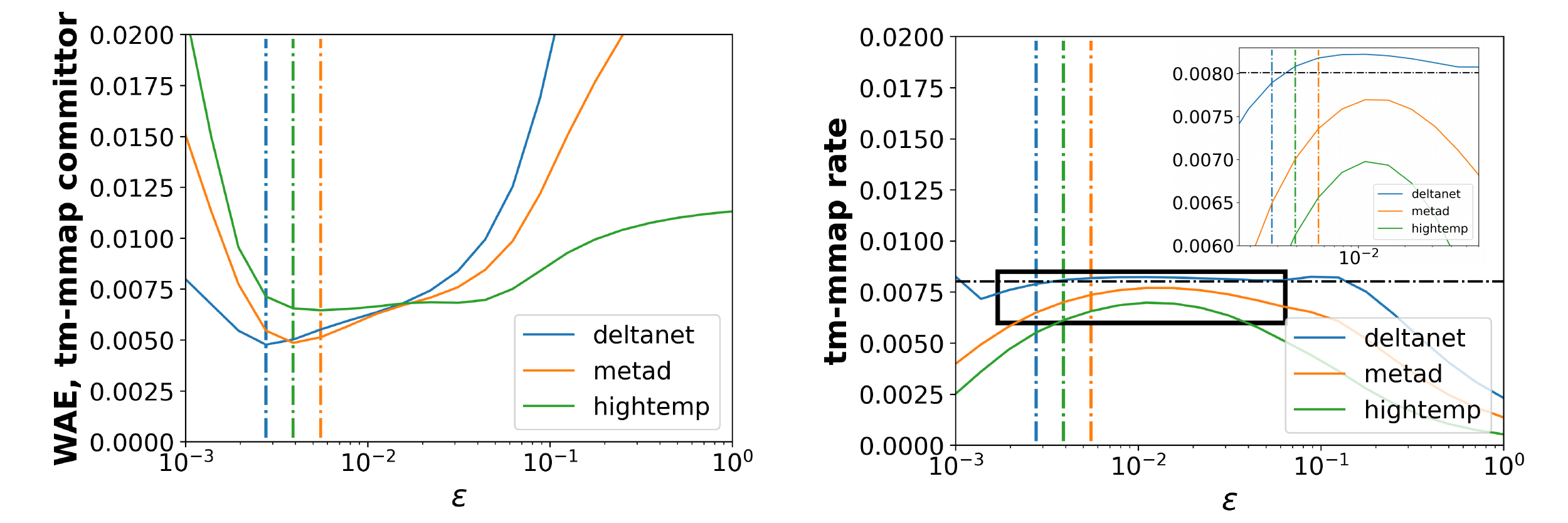}
  \caption{An application of the {\tt tm--mmap} to the Moro-Cardin system (Section \ref{sec:MoroCardin}). Left:
    Weighted absolute error \eqref{eqn:WAE}--\eqref{eqn:w} for the committor computed by {\tt tm-mmap} using the {\tt hightemp}, {\tt metad}, and {\tt deltanet} datasets. The committor computed using the finite element method is taken as the ground truth. The dotted lines indicate the epsilon chosen from the Ksum test (Algorithm \ref{alg:kdoublesum}).
    Right: A comparison of the transition rates computed from the {\tt tm-mmap} committors (colored curves) for a range of values of the bandwidth coefficient $\epsilon$ to the transition rate obtained via using the finite element method (dashed black line).}
\label{fig:morocardin_errors_rates}
\end{figure*}

The committors and the reactive currents for the {\tt hightemp}, {\tt metad} and {\tt deltanet} datasets are shown in Figure \ref{fig:morocardin_validation}. The errors in the committor and the transition rate measured against the FEM solutions for all three datasets are plotted in Fig. \ref{fig:morocardin_errors_rates}. The dashed vertical lines correspond to the values of the bandwidth parameter $\epsilon$ obtained using Agorithm \ref{alg:kdoublesum}. The error plots indicate that the {\tt deltanet} dataset leads to the most accurate results and most robust with respect to the choice of the bandwidth parameter $\epsilon$. In particular, the discrepancy between the transition rate computed by {\tt tm-mmap} for the {\tt deltanet} dataset and the FEM rate is within 5\% for a broad range of epsilon values: $2\cdot 10^{-3} < \epsilon < 10^{-1}$. The results obtained for the {\tt hightemp} dataset are the least accurate and least robust.

\subsection{Alanine dipeptide with two dihedral angles}
\label{sec:aladip2D}
\begin{figure}
\centering
    \includegraphics[width=0.45\textwidth]{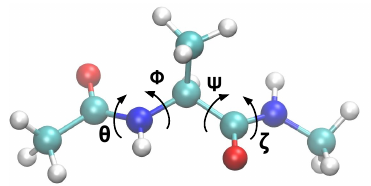}

\caption{Alanine dipeptide molecule with dihedral angles labelled.}
\label{fig:aladip}
\end{figure}

\begin{figure}
\centering
    \includegraphics[width=0.45\textwidth]{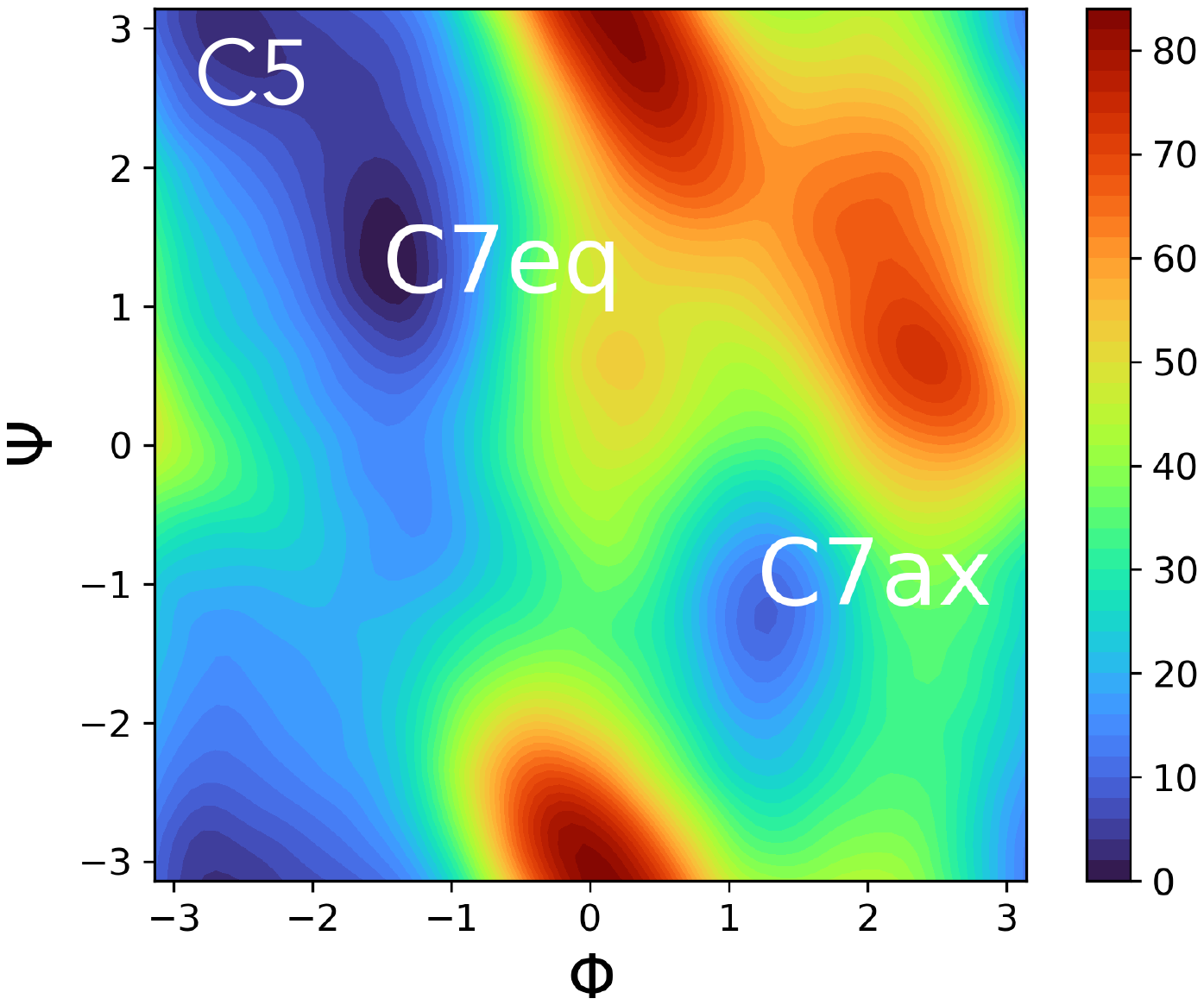}
\caption{Free energy landscape (kJ/mol) for alanine dipeptide in vacuum.}
\label{fig:aladipFE}
\end{figure}

Alanine dipeptide (Figure \ref{fig:aladip}) is a small biomolecule commonly used for benchmarking rare event algorithms for MD simulation. Its free energy in two dihedral angles $\phi$ and $\psi$ has three local minima (Figure \ref{fig:aladipFE}) making it a popular test example in chemical physics.~\cite{2006string,cameron2013estimation,plumed2,trstanova2020local} The close minima C5 and C7eq are separated from each other by a low free energy barrier of about  5 kJ/mol. The remote minimum C7ax is separated from C7eq by a barrier of approximately 40 kJ/mol.  The transition between C5 and C7eq was quantified using {\tt mmap}\cite{evans2021computing} because adequate data for this purpose could be collected by sampling from the invariant density. Now we are interested in the transition between the C7ax and C7eq metastable states. Robust enhanced sampling such as WTMETAD is required for obtaining an adequate data coverage for this transition.  

A typical set of collective variables effectively representing its motion consists of two to four of its dihedral angles.~\cite{2006string,plumed2,valsson2016,gao2021diffusion}
In contrast to previous uses of diffusion maps on alanine dipeptide~\cite{ferguson2010systematic, rohrdanz2011determination, trstanova2020local} we utilize the dihedral angles as input coordinates for diffusion maps. For numerical tests, we choose the set of only two dihedral angles $\phi$ and $\psi$, as the finite element method can be used for validation in this case. The space $\Omega$ of the collective variables $(\phi,\psi)$ is the square $[-\pi,\pi]^2$ glued into the torus $\mathbb{T}^2$. 

In order to generate datasets for alanine dipeptide, we ran a long metadynamics trajectory using GROMACS~\cite{van2005} and WTMETAD.
A velocity-rescaling termostat was used to set the temperature to 300K in a vacuum under constant number, volume, and temperature (NVT) conditions. The timestep was set to 2 femtoseconds, while the total simulated time interval was 10 nanoseconds. The parameters for WTMETAD were $h=1.2$, $\sigma=0.2$, and $\gamma = 6$. Then the dataset denoted by {\tt metad} was obtained by subsampling the trajectory data uniformly in time to make a total of $N=10^4$ points. The second dataset {\tt deltanet} obtained from the trajectory data by Algorithm \ref{alg:deltanet} with $\delta=0.04$ also contains $N = 10^4$ points.

\begin{figure}
\centering
    \includegraphics[width=0.45\textwidth]{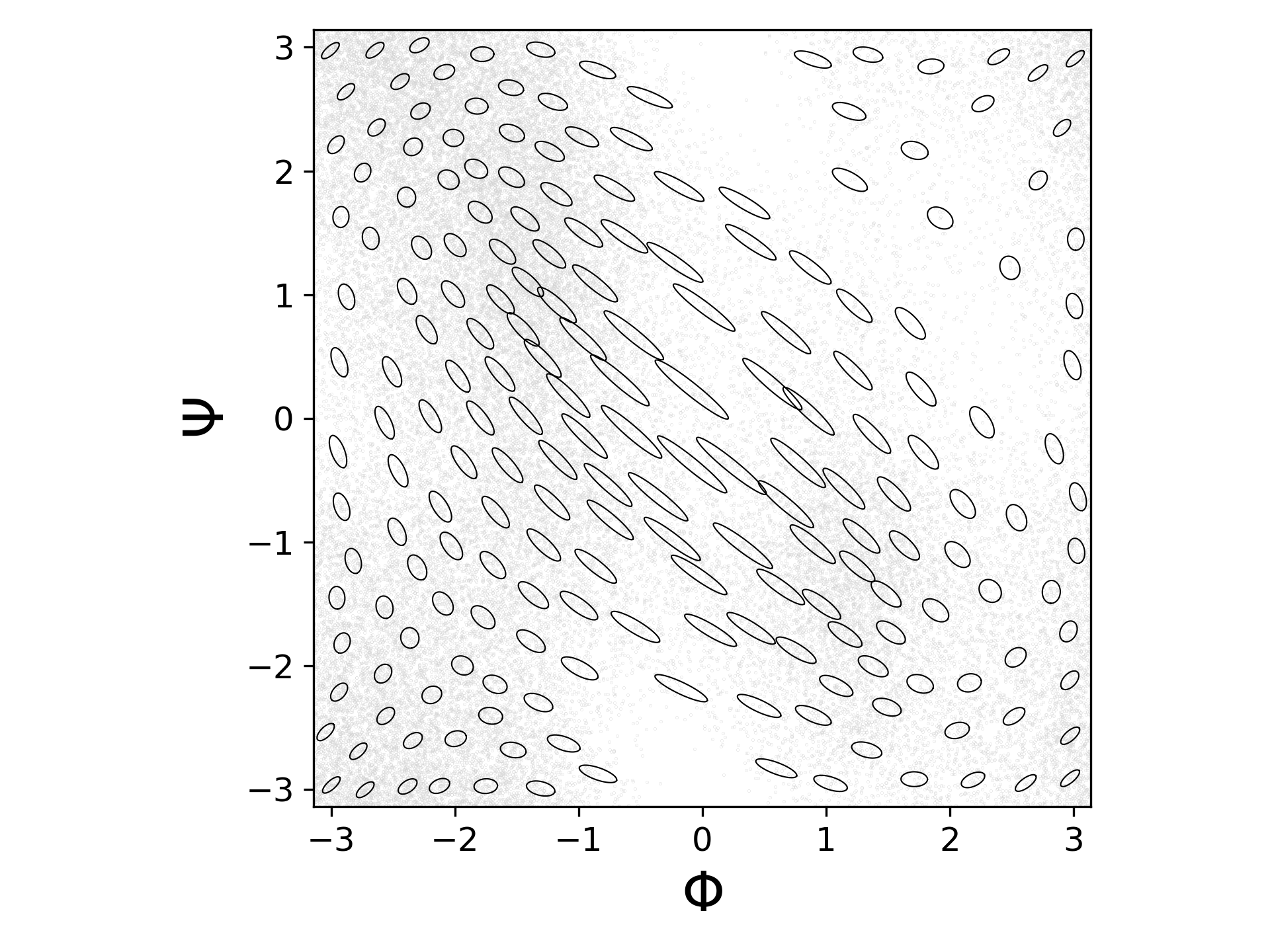}
  \caption{A depiction of the diffusion matrix for alanine dipeptide in two dihedral angles. The ellipses correspond to the principal components of the estimated diffusion matrices evaluated at a representative subset of metadynamics data (faint grey dots). }
  \label{fig:aladip2Ddiffusions}
\end{figure}

The diffusion matrices $M(x_i)$, $1\le i\le N$, are visualized in
Figure~\ref{fig:aladip2Ddiffusions}. 
We utilized the software PLUMED~\cite{plumed2} to obtain the derivatives of the dihedral angles with respect to the atomic coordinates and to obtain the restrained averages.
The reactant and product sets $A$ and $B$ are the circles of radius $0.35$ centered at the C7eq and C7ax minima located at $(-1.419, 1.056)$ and $(1.230, -1.206)$ respectively.

The free energy  $F$ in $x = (\phi, \psi)$ shown in Figure~\ref{fig:aladipFE} was obtained from a separate WTMETAD simulation with biasing on $\phi$ and $\psi$. The target measure is $\mu(x) = e^{-\beta F(x)}$.

\begin{figure*}
  \centering
   \includegraphics[width=\textwidth]{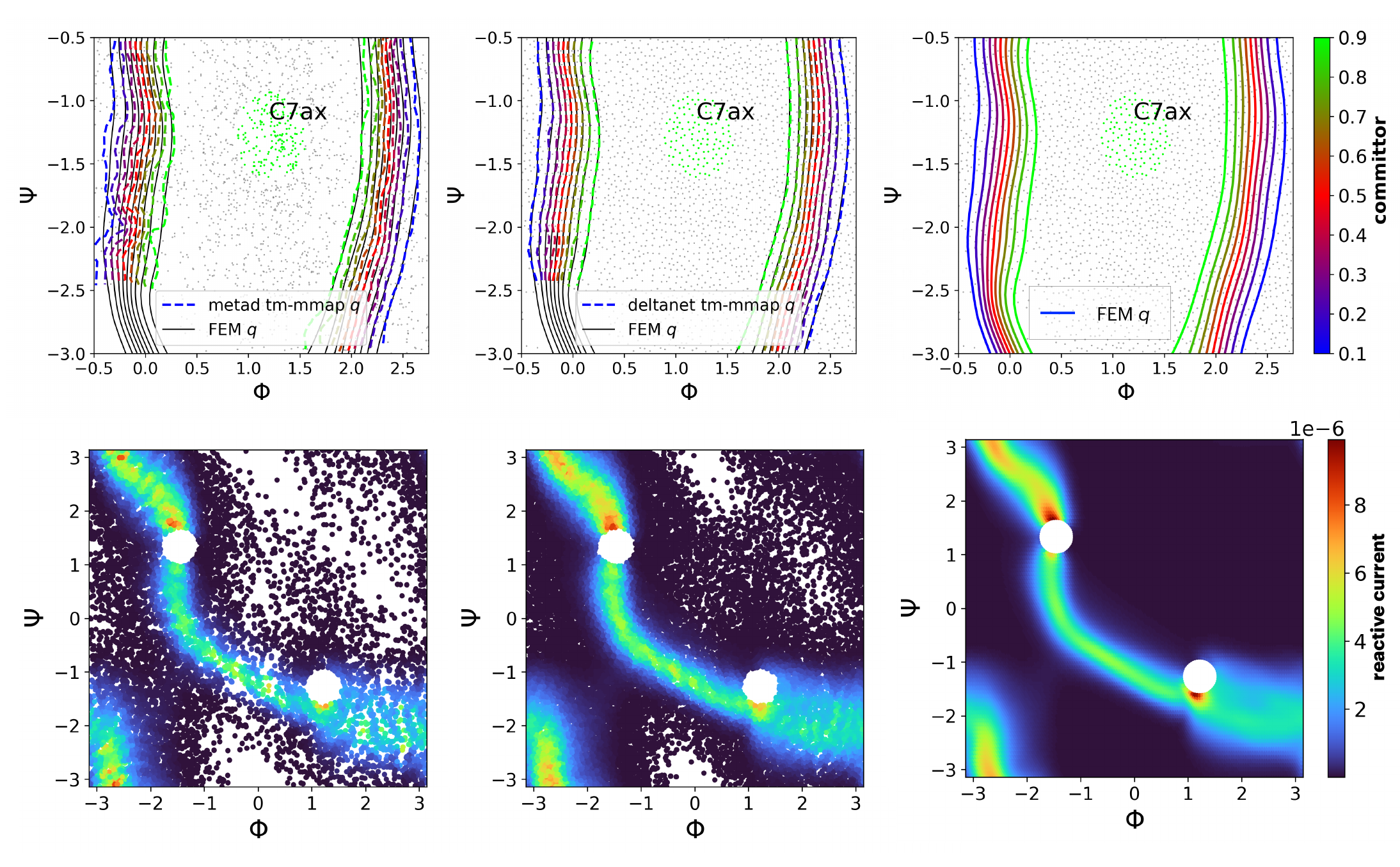}
  \caption{
  The level sets of the committor and magnitude of the reactive current computed for alanine dipeptide in vacuum in $(\phi, \psi)$ coordinates using {\tt tm-mmap}  for datasets {\tt metad} (left column), {\tt deltanet} (middle column), and using the finite element method (right column).
  }
  \label{fig:aladip_validation}
\end{figure*}

\begin{figure*}
     \centering
    \includegraphics[width=\textwidth]{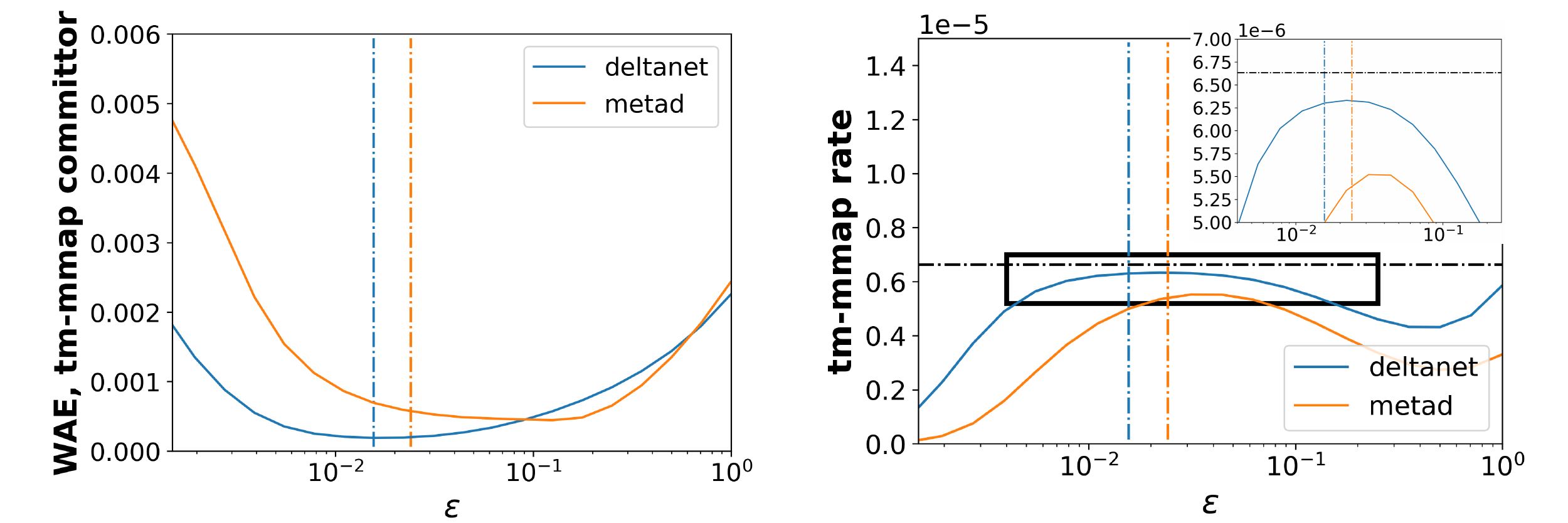}
  \caption{ Left: 
    Weighted absolute errors \eqref{eqn:WAE}--\eqref{eqn:w} of the committor computed by {\tt tm-mmap} for alanine dipeptide in vacuum in $(\phi, \psi)$ dihedral angles with respect to the committor computed by the finite element method. The dotted lines indicate the epsilon chosen by Algorithm \ref{alg:kdoublesum}.
    Right: Comparison of rates for
      {\tt tm-mmap} on a range of $\epsilon$ values, with zoom to the area of the closest rate approximation. The dotted black line indicates the rate obtained by the finite element method.}
  \label{fig:aladip_errors_rates}
\end{figure*}

The committor and the reactive current obtained using {\tt tm-mmap} on the datasets {\tt metad} and {\tt deltanet} are compared to those computed by the finite element method (FEM) in Figures \ref{fig:aladip_validation} and \ref{fig:aladip_errors_rates}. For both datasets, the committor and the current by {\tt tm-mmap} are similar to those by the FEM, but the results for {\tt deltanet} are considerably less noisy. Moreover, the weighted absolute error  \eqref{eqn:WAE}--\eqref{eqn:w} for the committor for {\tt deltanet} is smaller than that for {\tt metad} for most choices of $\epsilon$. In particular, it is smaller for the values of $\epsilon$ suggested by Algorithm \ref{alg:kdoublesum}. The transition rate approximated by {\tt tm-mmap} for the {\tt deltanet} dataset differs by less than 10\% from the rate obtained by FEM for a broad range of epsilon values. Notably, the $\epsilon$-value suggested by Algorithm \ref{alg:kdoublesum} is very close to optimal. The FEM transition rate is $6.6\cdot10^{-6}$ ps$^{-1}$, while the {\tt tm-mmap} with {\tt deltanet} transition rate at $\epsilon$ suggested by { the Ksum test} (Algorithm \ref{alg:kdoublesum} in Appendix \ref{sec:algorithms}) is $6.3\cdot 10^{-6}$ ps$^{-1}$.

\section{Application: alanine dipeptide with four dihedral angles}
\label{sec:aladip4D}
While only two dihedral angles are needed for enhanced sampling of the configurational space of alanine dipeptide~\cite{tiwary2015time}, at least three dihedral angles are required to adequately describe the transition mechanism from C7eq to C7ax.~\cite{2006string,bolhuis2000reaction,tiwary2013metadynamics,medhi2022,vani2022computing} The {\tt tm-mmap} algorithm is a suitable tool for computing the committor in 4D. Therefore, we consider the dynamics of alanine dipeptide represented by four dihedral angles $(\phi,\psi,\theta,\zeta)$. The collective variable space is the four-dimensional torus $\mathbb{T}^4$ obtained by an appropriate gluing of the four-dimensional hypercube $[-\pi,\pi]^4$. 

\subsection{Simulation details}
To generate the data, we first run WTMETAD to build a bias potential on a grid in the $\phi$-$\psi$ variables with the same parameters as
in Section~\ref{sec:aladip2D}. We then run a trajectory with the fixed bias to obtain $N=10^4$ data points $x_i \in\mathbb{T}^4$.  
Utilizing the kernel density estimate $\rho_{\epsilon}(x)$ from~\eqref{eqn:kde} and the stored metadynamics bias $U(x)$ we obtained an estimate to the target measure using equation \eqref{eqn:mu_estimate}.

\begin{figure*}
    \includegraphics[width=1\textwidth]{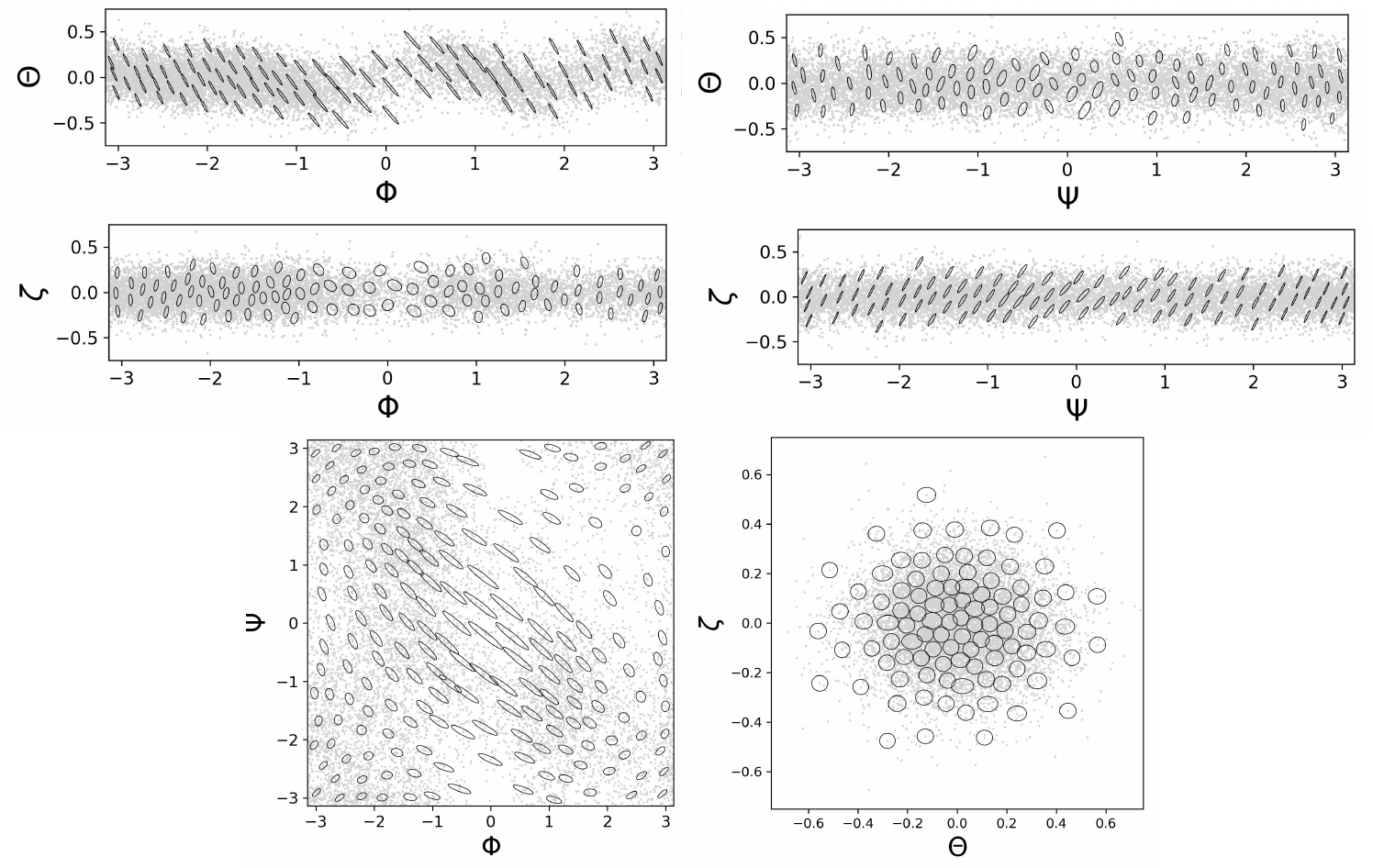}
    \caption{
    Approximations of the principal components for the $4 \times 4$ diffusion tensors obtained in all dihedrals for alanine dipeptide in vacuum. Each plot corresponds to principal components of the $2 \times 2$ submatrices associated to each pair of dihedral angles.}
    \label{fig:aladip_4DM}
\end{figure*}
We computed the $4 \times 4$ diffusion matrices $M(x_i)$ with the aid of the package PLUMED from equation~\eqref{eqn:diffusion_tensor_formula} as described in Ref.~\onlinecite{2006string} and Appendix A of Ref.~\onlinecite{evans2021computing}. The $2 \times 2$ sub-matrices corresponding to each possible pair of dihedral angles are visualized in Figure~\ref{fig:aladip_4DM}. The sub-matrices still correspond to a conditional average on $(\phi, \psi, \theta, \zeta)$ rather than a particular pair of dihedrals. Despite this, the $\phi$-$\psi$ sub-matrices are virtually indistinguishable from those in Figure~\ref{fig:aladip2Ddiffusions} for alanine dipeptide in only two dihedrals. This similarity to the un-averaged diffusion matrices in $\phi, \psi$  was also noted in Ref.~\onlinecite{branduardi2012}. 
We also point out the consistency of the
$\phi$-$\theta$  and $\psi$-$\zeta$ plots in Figure~\ref{fig:aladip_4DM} with the results obtained in Refs.~\onlinecite{tiwary2013metadynamics} and \onlinecite{ma2006dynamic}, respectively. These similarities indicate that the partial derivatives of each dihedral angle with respect to the atomic coordinates are approximately constant on any fiber corresponding to a fixed value of this angle. 

The reactant and product sets $A$ and $B$ in $(\phi, \psi, \theta, \zeta)$-coordinates were balls of radius 0.3 centered at $(-1.443, 1.282, 0.027, -0.075)$ and $(1.230, -1.206, -0.013,  0.099)$ respectively.

\subsection{Results}
\begin{figure*}
  \centering
    \includegraphics[width=\textwidth]{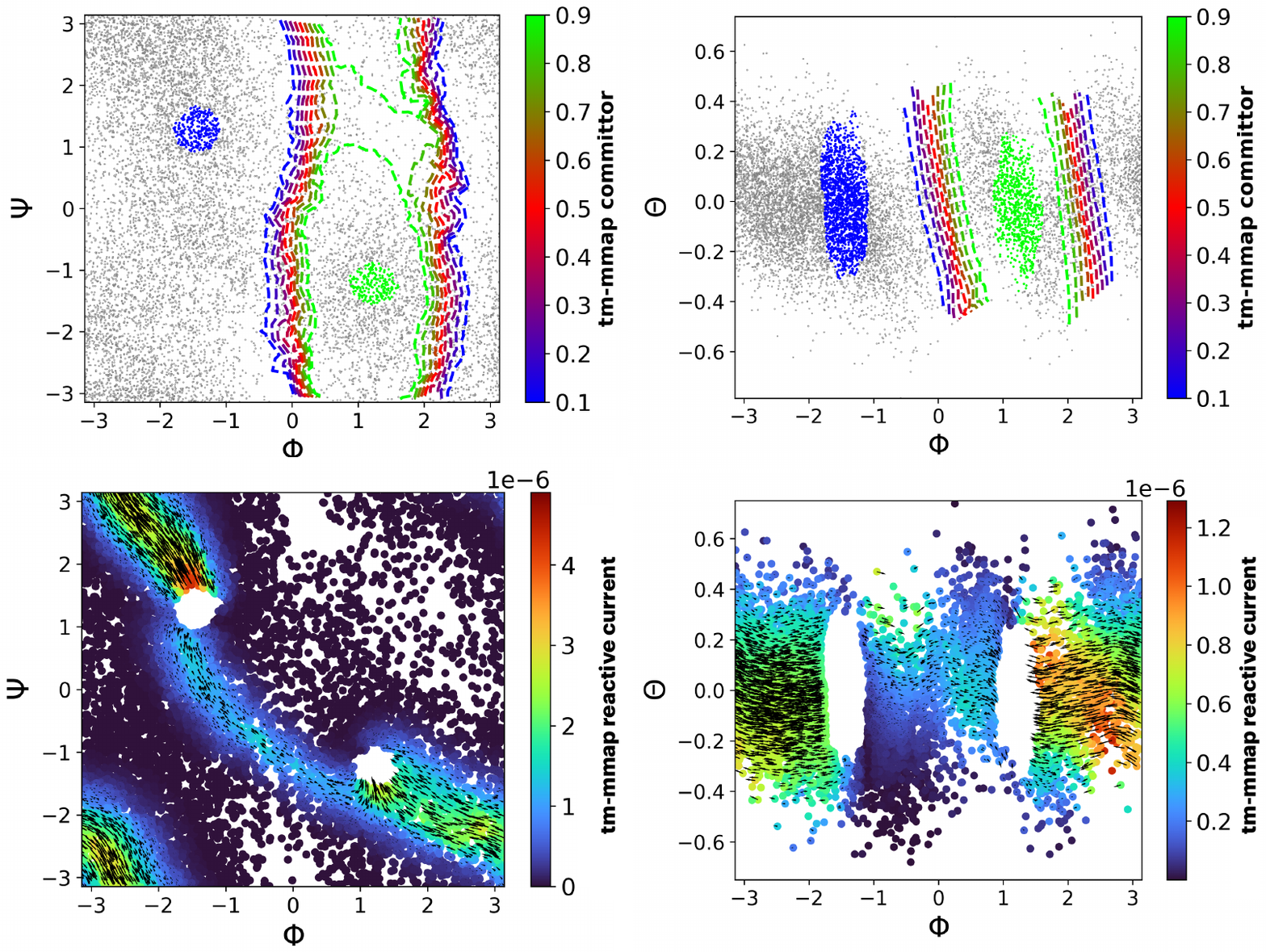}
  \caption{ The committor (top row) and the reactive current (bottom row) for alanine dipeptide in four dihedral angles computed using {\tt tm-mmap}. The smoothed projection onto the $(\phi,\psi)$-space (left column) and the $(\phi,\theta)$-space (right column) was computed using equation \eqref{eqn:smoothproj}.
 } 
  \label{fig:aladip_4var}
\end{figure*}
The committor and the reactive current computed using {\tt tm-mmap} are displayed in Figure~\ref{fig:aladip_4var}. In order to visualize functions of four variables in 2D we used projection and smoothing. For example, the committor  $\bar{q}(\phi,\psi)$ shown in Figure~\ref{fig:aladip_4var} (top left) is defined by:
\begin{align}
\bar{q}(\phi,\psi) &= \frac{1}{|S|}\sum_{x'\in S}q(x'),\quad{\rm where} \label{eqn:smoothproj}\\
S=\{x'&=(\phi',\psi',\theta',\zeta')
\in\mathbb{T}^4~| \notag \\
&\sqrt{(\phi-\phi')^2 + (\psi-\psi')^2} < r\}.\notag
\end{align}
The radius $r$ was chosen to be 0.2 by trial and error. The committor $\bar{q}(\phi,\theta)$ and the reactive currents displayed in Figure~\ref{fig:aladip_4var} were defined in a similar manner. Note that the reactive current in $(\phi,\theta)$ is comparable in magnitude to the reactive current in $(\phi,\psi)$.

The transition rate for alanine dipeptide in four dihedral angles computed using \eqref{eqn:rate_gamma_reweight} is $2.0\cdot 10^{-6}$ ps$^{-1}$. This rate is comparable with the one obtained in Ref.~\onlinecite{vani2022computing} as a direct estimate from a very long unbiased trajectory trajectory simulated for 2.5 microseconds: $1.4\cdot 10^{-6}$ps$^{-1}$.

\section{Discussion}
\label{sec:discussion}
Our results indicate that the methodology developed in this work is a robust and powerful tool for quantifying rare events in systems described in terms of collective variables and governed by SDE~\eqref{eqn:cvsde}. The dimensionality of the space of collective variables can be notably higher than what is feasible by traditional PDE solvers such as finite difference and finite element methods. The {\tt tm-mmap} algorithm for computing the committor proposed in this work requires three pieces of input: a set of data points, the diffusion matrix and the target measure {(up to a normalizing constant)} evaluated at the data points. Both the target measure and the diffusion matrix can be straightforwardly computed using standard software.  The fact that the data points can be sampled from an arbitrary density with respect to which the target measure is absolutely continuous makes {\tt tm-mmap} highly practical as any standard enhanced sampling software can be used for generating the input dataset. 

The computation of the diffusion matrices at the data points is the costliest part of the required input. For example, on a standard laptop a run of the {\tt tm-mmap} algorithm for alanine dipeptide using a dataset of size $N=10^4$ takes from 5 to 30 seconds (depending on sparsity) whereas the computation of the diffusion matrices for it in four dihedral angles takes about 4 hours. In general, other approaches such as the Kramer's-Moyal expansion~\cite{banisch2020,singer2008, nuske2021spectral} and related ``swarms of trajectories'' approaches~\cite{maragliano2014comparison, roux2021string} can be used, and computational cost can be reduced by computing diffusion matrices on a subset of the data.~\cite{kushnir2012anisotropic}

We would like to point out that since the density for the input dataset is arbitrary, the user can generate a large initial dataset and then prune it in any manner. Our results for the Moro-Cardin system and for alanine dipeptide in two dihedral angles suggest that spatially quasi-uniform datasets obtained by constructing delta-nets (Algorithm \ref{alg:deltanet}) lead to the best results. It is very important that the committor and the transition rate remain close to the FEM rate treated as the ground truth for a broad range on values of $\epsilon$. 

This observation suggests an investigation into how the numerical error in the committor computed by {\tt tm-mmap} depends on the density of the input dataset. Is the uniformly distributed input dataset optimal? We leave this investigation for future work.

An important feature of the proposed methodology is that it yields numerical values for the reactive current and the transition rate. Comparing the transition rate for alanine dipeptide represented in $(\phi,\psi)$ ($\nu_{AB} = 6.3\cdot 10^{-6}$ ps$^{-1}$)  and in $(\phi,\psi,\theta,\zeta)$ ($\nu_{AB} = 2.0\cdot 10^{-6}$ ps$^{-1}$) we observe that the former is a factor of $\sim 3.2$ larger than the latter. On the other hand, the rate for the four dihedral angles is in the ballpark of the estimate obtained from an unbiased very long trajectory.~\cite{vani2022computing} This suggests the deficiency of two-dimensional representation of alanine dipeptide and emphasizes the importance of the ability of {\tt tm-mmap} to work in higher dimensions. 

{ An important issue that has not been not addressed in this work but which needs to be addressed in future research is the size of the dataset that is sufficient to provide a desired accuracy.
This issue becomes especially paramount in connection with applying {\tt tm-mmap} in moderately higher dimensional CV spaces. As we have mentioned, the main limitation of {\tt tm-mmap} in its present version comes from the need to compute the diffusion tensor as all data points. It is of interest to assume that the dynamics of the system under consideration are intrinsically low-dimensional and extend diffusion maps to systems with degenerate diffusion.
}

\section{Conclusion}
\label{sec:conclusion}

We proposed a methodology for quantifying transition processes in molecular or biomolecular systems described in collective variables using the framework of the transition path theory and featuring the committor solver {\tt tm-mmap} (Algorithm \ref{alg:tmap}) based on diffusion maps and suitable for high dimensions (at least 4D as markedly shown in this work). The Mahalanobis kernel is capable of capturing the effects of variable and anisotropic diffusion ubiquitously arising in the dynamics in collective variables. The ``target measure'' variant of diffusion maps, on one hand, allows for input datasets being sampled from an arbitrary density, while on the other hand, requires the input of the target measure that is usually unknown a priori. This challenge has been overcome by proposing a simple technique described in Section \ref{sec:FE&M} that organically works with enhanced sampling and {\tt tm-mmap}. The summary of our technical contributions is the following.
\begin{itemize}
\item An adjustment of the target measure diffusion map algorithm~\cite{banisch2020} for the use with the Mahalanobis kernel \eqref{eqn:mahal_kernel} resulting in the {\tt tm-mmap} algorithm (Algorithm \ref{alg:tmap}). The key step is the right normalization of the kernel function \eqref{eqn:rnmkernel}.
\item A convergence proof of the discrete matrix operator built by {\tt tm-mmap} to the generator for the overdamped Langevin dynamics in collective variables \eqref{eqn:mgen}, given in Appendix \ref{sec:appendixA}.
\item A technique for estimating the unknown target measure in collective variable space that works with {\tt tm-mmap} and any enhanced sampling algorithm with a stored biasing potential. The key formula is \eqref{eqn:mu_estimate}.
\item The post-processing of the enhanced sampling data resulting in a spatially quasi-uniform delta-net (Algorithm \ref{alg:deltanet}).
\end{itemize}

The numerical tests conducted on the Moro-Cardin system and on alanine dipeptide with two dihedral angles demonstrated the accuracy of {\tt tm-mmap} for the committor, the reactive current, and the transition rate compared against the benchmark data produced by the finite element method. With regard to practical implementation, the tests against the benchmark illustrated the robustness of {\tt tm-mmap} to the choice of the bandwidth parameter $\epsilon$ and close to optimal
heuristic estimates for $\epsilon$ by the Ksum test (Algorithm \ref{alg:kdoublesum})~\cite{berry2015nonparametric,berry2016variable,giannakis2019data,davis2021graph}. The spatially quasi-uniform datasets lead to the most accurate and the most robust results.

The application of {\tt tm-mmap} to alanine dipeptide in four dihedral angles { has yielded} a transition rate estimate from C7eq to C7ax that is reasonably close to the one extracted from an extremely long trajectory.~\cite{vani2022computing}

We conclude that {\tt tm-mmap} is a promising tool for rare-event quantification. It is accurate, robust, suitable for high dimensions, and easy-to-use.

\begin{acknowledgments}
  This work was  partially supported by AFOSR MURI grant FA9550-20-1-0397 (MC)
and  NSF CAREER grant CHE-2044165 (PT). 
Computing resources Deepthought2, MARCC and XSEDE (project TG-CHE180053) were used to generate molecular simulation data. We thank Dr. Bodhi Vani for valuable discussion on MD simulations of alanine dipeptide and Mr. Zachary Smith for advice on reweighting with WTMETAD. We also thank the anonymous reviewers for their valuable comments that helped us to improve the presentation.
\end{acknowledgments}

\section*{Data Availability Statement}
Implementation of the diffusion map code was inspired by the approach in the {\tt pydiffmap} library.~\cite{pydiffmap} 
The code for {\tt tm-mmap} will be made available at \url{https://github.com/aevans1/targetmeasure-mmap}.

\appendix

\section{Convergence proof for {\tt tm-mmap}}
\label{sec:appendixA}
In the following, we state and prove the main result from Section~\ref{sec:tm-mmap}.
First, we recall the necessary definitions.
The Mahalanobis kernel is defined as:
\begin{equation}
	k_{\epsilon}(x,y) = \exp\left(-\frac{(x-y)^{\top}(M^{-1}(x) + M^{-1}(y))(x-y)}{4\epsilon}\right).
\end{equation}
The \textbf{kernel density estimator} is
\begin{equation}\label{eqn:mmap_density}
\rho_{\epsilon}(x):= \frac{1}{c_{\epsilon}(x)}\int_{\Omega} k_{\epsilon}(x, y) \rho(y) dy,
\end{equation}
where
\begin{equation}\label{eqn:normalizing_constant}
c_{\epsilon}(x):= {\int_{\Omega}k_{\epsilon}(x,y)dy = }
(2\pi\epsilon)^{d/2} |M|^{1/2}(x).
\end{equation}
{ The \textbf{right-normalized kernel} that neutralizes the influence of the sampling density $\rho$ and imposes the effect of the target measure $\mu$ is:} 
\begin{equation}\label{eqn:mu_weighted_mahal}
	k_{\epsilon, \mu}(x, y):= \frac{k_{\epsilon}(x, y) \big(|M|^{-1/2}(y) \mu(y)\big)^{1/2}}{\rho_{\epsilon}(y)}.
\end{equation}
The corresponding integral operator is defined as: 
\begin{equation}\label{eqn:mu_weighted_operator}
	\mathcal{K}_{\epsilon, \mu}f(x):= \int_{\Omega} k_{\epsilon,\mu}(x,y) f(y) \rho(y) dy.
\end{equation}
 { Then the {\bf Markov operator} or} the \textbf{left-normalized kernel operator} is:
\begin{equation}\label{eqn:mu_weighted_markov}
	\mathcal{P}_{\epsilon, \mu}f(x):= \frac{\mathcal{K}_{\epsilon, \mu}f(x)}{ \mathcal{K}_{\epsilon, \mu}\mathbf{1}(x)},
\end{equation}
where $\mathbf{1}(x) \equiv 1$ for all $x \in \Omega.$
Finally, the { {\bf generator operator}} is defined as:
\begin{equation}\label{eqn:mu_weighted_gen}
	\mathcal{L}_{\epsilon,\mu} f(x) := \frac{1}{\epsilon}\left(\mathcal{P}_{\epsilon, \mu}f(x) - f(x)\right).
\end{equation}
Our goal is to evaluate the limit as $\epsilon \to 0$:
\begin{equation}\label{eqn:limiting_generator}
	\lim\limits_{\epsilon \to 0}\mathcal{L}_{\epsilon,\mu} f(x).
\end{equation}

{ We will adopt two technical assumptions simplifying the calculations. The first assumption restricting the class of manifolds $\Omega$ is borrowed from Ref.~\onlinecite{evans2021computing}.}
\begin{assumption}
\label{Ass1}
The range of $x$ representing the set of collective variables constitutes a $d$-dimensional manifold $\Omega$ which is either $\mathbb{R}^d$, the $d$-dimensional torus $\mathbb{T}^d$, or a direct product of torus $\mathbb{T}^{k}$ and $\mathbb{R}^{d-k}$ . In all cases, $\Omega$ is of the form
\begin{equation}
\label{eqn:manifold}
\Omega =\mathbb{T}^{k}\times\mathbb{R}^{d-k} ,
\quad\text{for some}\quad 0\le k\le d.
\end{equation}
By the torus $\mathbb{T}^k$, $1\le k\le d$, we mean the ``flat" torus, i.e., the direct product of intervals with periodic boundary conditions. Therefore,
\begin{align*}
\mathbb{T}^k\times \mathbb{R}^{d-k} =& [a_1,b_1]\times\ldots\times[a_k,b_k]\times \mathbb{R}^{d-k}~~{\rm with}\\
&(x_1,\ldots,x_{l-1},a_l,x_{l+1},\ldots,x_d) \\
=& (x_1,\ldots,x_{l-1},b_l,x_{l+1},\ldots,x_d),\\
&1\le l\le k.
\end{align*}
The metric on such a torus is locally Euclidean \cite{nikulin1987geometries}, i.e., within any open ball of radius 
\begin{equation}
\label{eqn:Reuc}
R_{Euc}: = \min_{1\le l\le k}\frac{|b_l-a_l|}{2}.
\end{equation} 
Therefore, the metric on $\Omega$ is Euclidean if $\Omega=\mathbb{R}^d$ or locally Euclidean within any ball of radius $R_{Euc}$ if $\Omega =  \mathbb{T}^{k}\times\mathbb{R}^{d-k}$ for some $1\le k\le d$.\end{assumption}

{ The second assumption specifies the class of diffusion matrix $M(x)$.
\begin{assumption}
\label{Ass2}
The diffusion matrix-function $M(x)$ is symmetric positive definite for all $x\in\Omega$. Its inverse $M^{-1}(x)$ is a smooth matrix-function $M^{-1}:\Omega\rightarrow\R^{d\times d}$, the determinant of $M^{-1}(x)$ is bounded away from zero, and the entries of 
 $(M^{-1})_{ij}(x)$ as well as their first derivatives are $\frac{\partial (M^{-1})_{ij}(x)}{\partial x_{\ell}}$ are bounded.
\end{assumption}}

\begin{theorem}
\label{thm:main-theorem}
Suppose a manifold $\Omega$ and a diffusion matrix $M(x):\Omega\rightarrow\R^{d\times d}$ satisfy assumptions \ref{Ass1} and  \ref{Ass2} respectively.
Suppose we have a target measure, i.e., a nonnegative function $\mu: \Omega \to \R$, $\mu\in C^4(\Omega)$, such that $\int_{\Omega} \mu dx < \infty$. Suppose $\mu(x)$ is absolutely continuous with respect to the sampling density $\rho(x),$ 
and takes the form of a Gibbs measure $\mu(x) = e^{-\beta F(x)}$.
Then { for any smooth and bounded function $f:\Omega\rightarrow \mathbb{R}$,} the limit \eqref{eqn:limiting_generator} for the operator $\mathcal{L}_{\epsilon, \mu}$ constructed according to~\eqref{eqn:mmap_density},
\eqref{eqn:mu_weighted_mahal},
\eqref{eqn:mu_weighted_operator},
\eqref{eqn:mu_weighted_markov}, and 
\eqref{eqn:mu_weighted_gen} is
\begin{equation}
\label{eqn:main-eq}
\lim_{\epsilon\rightarrow 0} \mathcal{L}_{\epsilon,\mu} f(x) = \frac{\beta}{2}\mathcal{L}f(x)\quad\forall x\in\Omega,
\end{equation}
where 
\begin{equation}
\label{eqn:Lgen}
\mathcal{L}f = \big({\scriptsize -} M\nabla F + \beta\strut^{-1}\left(\nabla\cdot M\right)\big)^\top\nabla f + \beta\strut^{-1}{\sf tr}[M\nabla\nabla f]
\end{equation}
is the generator for the SDE 
\begin{align}
\label{eqn:main-sde}
dx_t =\big[-M(x_t)\nabla F(x_t) &+ \beta^{-1}\nabla\cdot M(x_t)\big]dt \nonumber\\
&+\sqrt{2\beta^{-1}}M^{1/2}(x_t)dw_t.
\end{align}
\end{theorem}

{ The proof of Theorem~\ref{thm:main-theorem} relies on the technical lemma that readily follows from Lemma C2 in Ref.~\onlinecite{evans2021computing} and calculations Appendix D in Ref.~\onlinecite{evans2021computing}.}
\begin{lemma}
\label{thm:lem2}
Let $\Omega$ be a manifold satisfying Assumption \ref{Ass1} and $G_{\epsilon}$ be an integral operator defined by
\begin{equation}
\label{eqn:Ge}
G_{\epsilon}f(x) = \int_{\Omega} e^{-\frac{1}{4\epsilon}(x-y)^\top [M^{-1}(x) +M^{-1}(y)](x-y)}f(y) dy,
\end{equation}
{ where the matrix-function $M$ satisfies Assumption \ref{Ass2} and the function $f:\Omega\rightarrow\mathbb{R}$ is smooth and bounded.}
Then
\begin{align}\label{eqn:kernel_exp}
G_{\epsilon}f(x) &=c_{\epsilon}(x)\left(f(x) +\epsilon\left[- \nabla f(x)^\top \omega_1(x) -f(x)\omega_2(x) \right.\right.\notag\\
&\left.\left.+\frac{1}{2}{\sf tr}\left[M(x)\nabla\nabla f(x)\right]\right] + O(\epsilon^2)\right),
\end{align}
where {$c_\epsilon(x)$ is defined in \eqref{eqn:normalizing_constant}, the vector function $\omega_1:\Omega\rightarrow\mathbb{R}^d$ is given by
\begin{equation}
\label{omega1}
\omega_1 = \frac{1}{2}\left( M \nabla \log |M^{-1/2}|(x) - \nabla \cdot M(x)\right),
\end{equation}
and the function $\omega_2:\Omega\rightarrow\mathbb{R}$ is an $O(1)$ function that depends on $M^{-1}(x)$ and the first and second derivatives at of its entries evaluated at $x$.
}  

\end{lemma}
As we prove Theorem \ref{thm:main-theorem}, the terms containing the function $\omega_2$ will cancel out, while the terms involving $\omega_1$ will contribute to the final result. 


\begin{proof}[Proof of Theorem 1.]
The expansion of $G_{\epsilon}f(x)$ in $\epsilon$ can be written as
	\begin{equation}\label{eqn:expansion_mahal}
		G_{\epsilon}f(x) = c_{\epsilon}(x_i) \left(f(x) + \epsilon Qf(x) + \mathcal{O}(\epsilon^{2})\right),
	\end{equation}
	where the first order terms in $\epsilon$ are collected in the operator $Q$ defined by 
	\begin{flalign}\label{eqn:Q}
	Qf &= -\omega_2f + \nonumber \\
	&\frac{1}{2}\Bigg[
	\Big(M \nabla \log |M^{1/2}| + \nabla\cdot M \Big)^\top\nabla f
	+ \tr[M\nabla \nabla f] 
	\Bigg]. &&
	\end{flalign}

To compute the Taylor expansion of the Markov operator $\mathcal{P}_{\epsilon, \mu}(x)$ defined in \eqref{eqn:mu_weighted_markov}, we will need the Taylor expansion of $\rho^{-1}_{\epsilon}(x)$ in $\epsilon$ {that enters the construction in \eqref{eqn:mu_weighted_mahal}.}
Applying Lemma \ref{thm:lem2} to $f(x) = \rho(x)$ and using the notation $Q$ we get:
\begin{equation}\label{eqn:rho_inv_mahal}
	\rho_{\epsilon}(x) = \rho(x) \left(1 + \epsilon \frac{Q \rho(x)}{\rho(x)} + \mathcal{O}(\epsilon^2)\right).
\end{equation}
Hence,
\begin{equation}\label{eqn:rho_inv_mahal}
	\rho_{\epsilon}^{-1}(x) = \rho^{-1}(x) \left(1 - \epsilon \frac{Q \rho(x)}{\rho(x)} + \mathcal{O}(\epsilon^2)\right).
\end{equation}

We now proceed to derive an asymptotic expansion of $\mathcal{P}_{\epsilon,\mu}f(x)$ in $\epsilon.$
To do so, we first introduce the following simplifying notation:
\begin{notation}
Given the target measure $\mu$ and diffusion tensor $M(x)$ with determinant $|M(x)|,$ we define \[\tilde{\mu}(x):= |M^{-1/2}(x)| \mu(x).\]
\end{notation}

Then, the numerator of the left-normalized kernel $\mathcal{P}_{\epsilon,\mu}$ from~\eqref{eqn:mu_weighted_markov} can be written as
\begin{equation}\label{eqn:numerator}
	\hspace{-0.25cm} \mathcal{K}_{\epsilon, \mu}f(x):= \int_{\Omega} k_{\epsilon}(x,y) f(y) \tilde{\mu}^{1/2}(y)\frac{\rho(y)}{\rho_{\epsilon}(y)} dy.
\end{equation}
From equation~\eqref{eqn:rho_inv_mahal} we calculate
\begin{equation}\label{eqn:divide_density_mahal}
	\frac{\rho(y)}{\rho_{\epsilon}(y)} 
	= 1 - \epsilon \frac{Q \rho(y)}{\rho(y)} + \mathcal{O}(\epsilon^{2}).
\end{equation}
Plugging equation~\eqref{eqn:divide_density_mahal} into the integrand of equation~\eqref{eqn:numerator}, utilizing
Lemma~\ref{thm:lem2}, and collecting higher order terms of $\epsilon$ we get:
\begin{align}\label{eqn:k_exp}
	\mathcal{K}_{\epsilon, \mu}f(x)
	  = \int_{\Omega}& \Bigg[ k_{\epsilon}(x, y)f(y)\tilde{\mu}^{1/2}(y) \nonumber \\
	  &\ \left(1 - \epsilon \frac{Q \rho(y)}{\rho(y)} + \mathcal{O}(\epsilon^2)\right)\Bigg] dy. 
\end{align}
Plugging the kernel expansion~\eqref{eqn:expansion_mahal} into \eqref{eqn:k_exp}  we obtain:
\begin{flalign}
	&\mathcal{K}_{\epsilon, \mu}f(x) \nonumber \\
	  &= \int_{\R^d} \Bigg[ k_{\epsilon}(x, y)f(y)\tilde{\mu}^{1/2}(y)\left(1 - \epsilon \frac{Q \rho(y)}{\rho(y)} + \mathcal{O}(\epsilon^2)\right)\Bigg] dy  \nonumber \\
	&= c_{\epsilon}\Bigg[f \tilde{\mu}^{1/2} {\scriptsize +} \epsilon\bigg(Q \big(f\tilde{\mu}^{1/2} \big) {\scriptsize -} f\tilde{\mu}^{1/2}\frac{Q  \rho}{\rho}\bigg)
    {\scriptsize +} \mathcal{O}(\epsilon^{2})\Bigg]. 
\end{flalign}

Similarly,
\begin{flalign}
	&\mathcal{K}_{\epsilon, \mu}\mathbf{1}(x) \nonumber \\
	  &= \int_{\R^d} \Bigg[ k_{\epsilon}(x, y)\tilde{\mu}^{1/2}(y)\left(1 - \epsilon \frac{Q \rho(y)}{\rho(y)} + \mathcal{O}(\epsilon^2)\right)\Bigg] dy \nonumber \\
	&= c_{\epsilon}\Bigg[\tilde{\mu}^{1/2} + \epsilon\bigg(Q \big(\tilde{\mu}^{1/2} \big) - \tilde{\mu}^{1/2}\frac{Q  \rho}{\rho}\bigg)
    + \mathcal{O}(\epsilon^{2})\Bigg]. 
\end{flalign}

For brevity in notation, we will omit the argument in $x$ from the remainder of the proof. 
Collecting powers of $\epsilon,$ we derive the operator $ \mathcal{P}_{\epsilon,\mu }$ from~\eqref{eqn:mu_weighted_markov} as
\begin{align}\label{eqn:markov_exp}
	\mathcal{P}_{\epsilon,\mu}f & =                                   \frac{f \tilde{\mu}^{1/2} + \epsilon\left(Q (f \tilde{\mu}^{1/2}) 
	- f\tilde{\mu}^{1/2}\frac{Q  \rho}{\rho}\right) 
	+ \mathcal{O}(\epsilon^{2})}
	{\tilde{\mu}^{1/2} + \epsilon\left(Q ( \tilde{\mu}^{1/2}) 
	- \tilde{\mu}^{1/2}\frac{Q  \rho}{\rho}\right) 
	+ \mathcal{O}(\epsilon^{2})} \nonumber \\                         
	& =\frac{f + \epsilon\left(\frac{Q (f \tilde{\mu}^{1/2})}{\tilde{\mu}^{1/2}} 
	- f\frac{Q  \rho}{\rho}\right) + \mathcal{O}(\epsilon^{2})}
	{1 + \epsilon\left(\frac{Q ( \tilde{\mu}^{1/2})}{\tilde{\mu}^{1/2}} 
	- \frac{Q  \rho}{\rho}\right) 
	+ \mathcal{O}(\epsilon^{2})} \nonumber \\
	& = f + \epsilon\left(\frac{Q (f \tilde{\mu}^{1/2})}{\tilde{\mu}^{1/2}}      
	- f\frac{Q (\tilde{\mu}^{1/2})}{\tilde{\mu}^{1/2}}\right) + \mathcal{O}(\epsilon^2).
\end{align}

We now will simplify the order $\epsilon$ term in~\eqref{eqn:markov_exp} using the definition of the operator $Q$ from~\eqref{eqn:Q}. { A building block for this calculation is an observation that for any  twice continuously differentiable functions $g,h:\Omega\rightarrow \mathbb{R}$ and a symmetric matrix function 
$M$ we have:}
\begin{flalign}
\tr(M\nabla\nabla(gh)) &= \sum\limits_{ij}M_{ij} \frac{\partial^2 (gh) }{\partial x_i \partial x_j} \nonumber \\
&= g \tr(M \nabla \nabla h) + h \tr(M \nabla \nabla g) \nonumber \\
&\qquad + 2 \nabla g^{\top} M \nabla h. \end{flalign}
Therefore,
\begin{align}
Q (gh) = -\omega_2 gh &+ \frac{1}{2}(h\nabla g + g\nabla h)^{\top}\Big(M \nabla \log |M|^{1/2} + \nabla \cdot M  \Big) \nonumber \\
&+ \frac{1}{2}\Big(g\tr(M\nabla\nabla h) +  h\tr(M\nabla\nabla g)\Big) \nonumber \\
&+ \nabla g^{\top} M \nabla h.
\end{align}
{ Hence, the order $\epsilon$ term of~\eqref{eqn:markov_exp} can be decoded as:}
\begin{align*}
\frac{Q (gh)}{h} &- g\frac{Q (h)}{h} \\
&= \frac{1}{2}\Big(M \nabla \log |M|^{1/2} +\nabla \cdot M \Big)^{\top} \nabla g \nonumber \\
&\qquad + \frac{\nabla g^{\top} M \nabla h}{h} +  \frac{1}{2}\tr(M\nabla\nabla g)  \\
	&= \frac{1}{2}\Bigg[\Big(M \nabla \log |M|^{1/2} + 2M \frac{\nabla h}{h} + \nabla \cdot M \Big)^{\top} \nabla g  \\
	&\qquad + \tr(M\nabla\nabla g) \Bigg].\\
\end{align*}
Note that the terms containing the function $\omega_2$ get cancelled. Substituting $g = f$ and $h = \tilde{\mu}^{1/2}$ into the last expression and plugging it into \eqref{eqn:markov_exp} we obtain:
\begin{align}
&(\mathcal{P}_{\epsilon,\mu}f - f)(x) \nonumber \\ &\qquad = 
	\frac{\epsilon}{2}\Bigg[
\Big(M \nabla \log |M|^{1/2} + 2M \nabla \log(\tilde{\mu}^{\frac{1}{2}}) + \nabla \cdot M  \Big)^{\top} \nabla f \nonumber\\
&\qquad \qquad + \tr(M\nabla\nabla f)\Bigg] + \mathcal{O}(\epsilon^2) \nonumber \\
&\qquad=  
	\frac{\epsilon}{2}\Bigg[
\Big(-M \nabla \log \mu + \nabla \cdot M  \Big)^{\top} \nabla f \nonumber \\
&\qquad \qquad + \tr(M\nabla\nabla f)\Bigg]  + \mathcal{O}(\epsilon^2).
\end{align}

Therefore, for $\mu(x) = e^{-\beta F(x)}$, the operator $\mathcal{L}_{\epsilon,\mu}$ is:
\begin{align}
	\mathcal{L}_{\epsilon,\mu}f(x) &= \frac{1}{\epsilon} (\mathcal{P}_{\epsilon, \mu}f - f)(x) \nonumber \\
	&= \frac{\beta}{2}\Big[\big(-M \nabla F(x) + \nabla \cdot M)^{\top}\nabla f \nonumber \\
	&\qquad\quad +  \beta^{-1}\tr(M\nabla \nabla f)\Big] + \mathcal{O}(\epsilon)  \nonumber \\
	&= \frac{\beta}{2} \mathcal{L}f + \mathcal{O}(\epsilon).
\end{align}
Thus, taking the limit as $\epsilon \to 0$ we obtain the desired result: 
\begin{equation}\label{eqn:limiting_generator_proof}
	\lim\limits_{\epsilon \to 0}\mathcal{L}_{\epsilon,\mu} f(x) = \frac{\beta}{2}\mathcal{L}f.
\end{equation}
\end{proof}


\section{Supplementary Algorithms}
\label{sec:algorithms}

Below we provide two key algorithms from the literature we have utilized for our numerical experiments. Algorithm~\ref{alg:kdoublesum} is the pseudocode for the procedure used to select the bandwidth $\epsilon$ in {\tt tm-mmap}, { the Ksum test}. Algorithm~\ref{alg:deltanet} is the pseudocode for obtaining { a delta-net}, i.e., a spatially quasi-uniform subsampling of the data. 
\begin{algorithm}[H]
  \SetAlgoNoLine
  \DontPrintSemicolon
  \KwIn{data $X = \{x_i\}_{i=1}^N$, diffusion matrices $\{M(x_i)\}_{i=1}^N,$ target measure $\mu$}
  \KwOut{bandwidth $\epsilon^{*}$}
  Choose a range of epsilon values $\mathcal{E}$, for example $\epsilon_i= 2^{i},$ $i=-20, -19,\ldots, 9, 10$ \;
  \For{$\epsilon$ in $\mathcal{E}$}{
    Compute the kernel $[K_{\epsilon}]$ \;
    Compute $\frac{\partial \log S(\epsilon)}{\partial \log \epsilon}$ according to~\eqref{eqn:logS_formula} \;
  }
  Choose $\epsilon^{*} = \argmax\limits_{\epsilon} \frac{\partial \log S(\epsilon)}{\partial \log \epsilon}.$
  \caption{Kernel Double Sum Test (Refs.~\onlinecite{berry2015nonparametric,berry2016variable,giannakis2019data,davis2021graph}) \label{alg:kdoublesum}}
\end{algorithm}

\begin{algorithm}[H]
  \SetAlgoNoLine
  \DontPrintSemicolon
  \KwIn{data $X = \{x_i\}_{i=1}^N$, spatial parameter $\delta$}
  \KwOut{delta-net $Z$}
  Choose initial ordering of the data set $\{x_i\}_{i=1}^N.$ \;
  Initialize $\delta$-net as $Z = \{x_1\}$ \;
  \For{$i = 2$ to $N$}{
    \If{$\min_{z \in Z}d(x_i, z) > \delta$}{
      $Z := Z \cup \{x_i\}$}
  }

  (Optional) Prune the delta-net for isolated points \;
  \For{$z \in Z$}{
    \If{$\min_{z' \in Z}d(z', z) > 2\delta$}{
      $Z := Z \backslash \{z\}$}
  }
  \caption{Delta-net (Ref.~\onlinecite{crosskey2017atlas}) \label{alg:deltanet}}
\end{algorithm}

\end{document}